\documentstyle[12pt]{article}
\input epsf
\parskip=\medskipamount
\def\al{\alpha}

   \def\Ga{\Gamma}
\def\de{\delta}   
\def\ep{\epsilon}
\def\la{\lambda}  
\def\te{\theta}   
\def\om{\omega}   \def\Om{\Omega}
\def\p{\phi}     %\def\p{\varphi}
\def\IC{\relax{\rm l\kern-.50 em C}}
\def\IE{\relax{\rm l\kern-.12 em E}}
\def\IK{\relax{\rm l\kern-.18 em K}}
\def\IL{\relax{\rm I\kern-.18 em L}}
\def\IN{\relax{\rm I\kern-.18 em N}}
\def\IR{\relax{\rm I\kern-.18 em R}}
\def\ii{\rm i\,}
\font\tenfrak=eufm10  \font\sevenfrak=eufm7  \font\fivefrak=eufm5
\newfam\frakfam
\textfont\frakfam=\tenfrak
\scriptfont\frakfam=\sevenfrak
\scriptscriptfont\frakfam=\fivefrak

\newtheorem{proposicion}{Proposition}
\def\Im{\mathop{\rm Im}\nolimits}

\def\frac#1#2{{#1\over #2}}
\def\fracpd#1#2{\frac{\partial #1}{\partial #2}}
\def\Cos{\mathop{\rm C}\nolimits}    % funcion Coseno
\def\Sin{\mathop{\rm S}\nolimits}    % funcion Seno
\def\Tan{\mathop{\rm T}\nolimits}    % funcion Tangente
\def\k{\kappa}                       % kappa
\begin{document}

\title{A non-linear Oscillator with quasi-Harmonic behaviour: \\
two- and $n$-dimensional Oscillators }

\author{
Jos\'e F. Cari\~nena$\dagger\,^{a)}$,
Manuel F. Ra\~nada$\dagger\,^{b)}$,
Mariano Santander$\ddagger\,^{c)}$ \\ and
Murugaian Senthilvelan$\sharp\,^{d)}$\\[4pt]
$\dagger$
  {\sl Departamento de F\'{\i}sica Te\'orica, Facultad de Ciencias} \\
  {\sl Universidad de Zaragoza, 50009 Zaragoza, Spain}  \\
$\ddagger$
  {\sl Departamento de F\'{\i}sica Te\'orica, Facultad de Ciencias} \\
  {\sl Universidad de Valladolid,  47011 Valladolid, Spain} \\
$\sharp$
  {\sl Centre for Nonlinear Dynamics, Bharathidasan University} \\
  {\sl Tiruchirapalli 620 024, India}
}
\maketitle
\date{}

\begin{abstract}
A nonlinear two-dimensional system is studied by making use of
both the Lagrangian and the Hamiltonian formalisms.
The present model is obtained as a two-dimensional version of a
one-dimensional oscillator previously studied at the classical
and also at the quantum level.
First, it is proved that it is a super-integrable system, and
then the nonlinear equations are solved and the solutions are
explicitly obtained.
All the bounded motions are quasiperiodic oscillations and the
unbounded (scattering) motions are represented by hyperbolic functions.
In the second part the system is generalized to the case of
$n$ degrees of freedom.
Finally, the relation of this nonlinear system with the harmonic
oscillator on spaces of constant curvature, two-dimensional
sphere $S^2$ and hyperbolic plane $H^2$, is discussed.
\end{abstract}

\begin{quote}
%%% {\it PACS codes:}
%%% {\enskip}02.30.Hq, {\enskip}02.40.Ky, {\enskip}45.20.JJ

\end{quote}

%%% 37J35 (2000-now) Completely integrable systems,
%%%    topological structure of phase space, integration methods
%%% 34A34 (1980-now) Nonlinear equations and systems, general
%%% 34C15 (1973-now) Nonlinear oscillations, coupled oscillators
%%% 70H06 (2000-now) Completely integrable systems and methods of integration
%%% 70K75 (2000-now) Nonlinear modes
{\vfill}

\footnoterule
{\noindent\small
$^{a)}${\it E-mail address:} {jfc@unizar.es} {\hskip44pt}
$^{b)}${\it E-mail address:} {mfran@unizar.es} \\
$^{c)}${\it E-mail address:} {santander@fta.uva.es} {\enskip}
$^{d)}${\it E-mail address:} {senthilvelan@cnld.bdu.ac.in} }
\newpage

%-----------------------------------------------------------
%%%%(Section 1.)
\section{Introduction and main results}

  Mathews and Lakshmanan studied in 1974 \cite{MaL74},\cite{LaRa03},
the equation
\begin{equation}
  (1 +\la x^2)\,\ddot{x} - (\la x)\,\dot{x}^2 + \al^2\,x  = 0
  \,,\quad\la>0\,,
\label{eq1}\end{equation}
as an example of a non-linear oscillator (notice $\al^2$ was written 
just as $\al$
in the original paper). In fact they considered (\ref{eq1})  as a 
particular case of
the  differential equation
$$
  y'' + f(y) y'^2 + g(y)   = 0 \,,
$$
that can be solved by using a two steps procedure:
(i) first a reduction of order can be fulfilled by the change $y'=p$,
$y''=p p'$, $p'=dp/dy$; (ii) then the corresponding first order equation
can be solved by using $\mu(y)= \exp\{2\int f(y)\,dy\}$, as an 
integrating factor.
In this case we have $f(x)=-\,\la\,x/(1 +\la x^2)$ and hence
$\mu(x)=1/(1+\la\,x^2)$; the general solution takes the form
$$
  x  = A \sin(\om\,t + \phi) \,,
$$
with the following additional restriction linking frequency and amplitude
$$
  \om^2  = \frac{\al^2}{1 + \la\,A^2} \,.
$$
That is, the equation (\ref{eq1}) represents a non-linear oscillator with
periodic solutions that they qualify as having a ``simple harmonic form".
The authors also proved that (\ref{eq1}) is obtainable from the Lagrangian
\begin{equation}
  L  = \frac{1}{2}\,\Bigl(\frac{1}{1 + \la\,x^2} \Bigr)\,(\dot{x}^2 - 
\al^2\,x^2)
\label{Lagn1}\end{equation}
which they considered as the one-dimensional analogue of the
Lagrangian density
\begin{equation}
  L  = \frac{1}{2}\,\Bigl(\frac{1}{1 + \la\,\phi^2} \Bigr)\,
    (\partial_{\mu}\phi\,\partial^{\mu}\phi - m^2\,\phi^2)    \,,
\end{equation}
appearing in some models of quantum field theory \cite{DeSS69},\cite{NiW72}.

The equation (\ref{eq1}) is therefore an interesting example of a
system with nonlinear oscillations with a frequency (or period) showing
amplitude dependence.
As a quantum system, the one-dimensional Schroedinger equation
involving the potential $x^2/(1 + g x^2)$ was considered in
\cite{BiDS73} as an example of anharmonic oscillator, and later
on studied in \cite{MaL75}-\cite{Fl81}; the three-dimensional quantum
problem  was considered in \cite{LaE75},\cite{BoV90}
(e.g., in \cite{LaE75} the Bohr-Sommerfeld quantization procedure was applied
in relation with some previous studies \cite{LiL70},\cite{VeW71}, in 
nonpolynomial
quantum mechanical models).
We observe that this system can also be considered as an oscillator
with a position-dependent effective mass
(see \cite{Le95} and references therein).

   The main objective of this article is to develop a deeper analysis
of the equation (\ref{eq1}) and the Lagrangian (\ref{Lagn1}), first
proving that this particular $\la$-dependent nonlinear system can be
generalized to the two-dimensional case, and even to the $n$-dimensional
case, and second, pointing towards a simple geometric interpretation of
the system so obtained.
In more detail, the plan of the article is as follows:
Sec.\ 2 is devoted to the properties of the $\la$-dependent kinetic part
of the Lagrangian and the $\la$-dependent two-dimensional free motion.
In Sec.\ 3, that must be considered as the central part of this article,
we study the $\la$-dependent two-dimensional oscillator;
we have divided this section in three parts:
in the first part we discuss the existence of Noether symmetries
for $\la$-dependent Lagrangians of the form
$$
  L(x,y,v_x,v_y;\la) = \frac{1}{2}\,\Bigl(\frac{1}{1 + \la\,r^2} \Bigr)
  \Bigl[\,v_x^2 + v_y^2 + \la\,(x v_y - y v_x)^2 \,\Bigr]  -  V(x,y;\la)\,.
$$
In the second part we discuss the properties of the $\la$-oscillator
described by the Lagrangian
$$
  L = \frac{1}{2}\,\Bigl(\frac{1}{1 + \la\,r^2} \Bigr)
  \Bigl[\,v_x^2 + v_y^2 + \la\,(x v_y - y v_x)^2 \,\Bigr]
  - \frac{\al^2}{2}\Bigl(\frac{r^2}{1 + \la\,r^2} \Bigr)\,,\quad
  r^2 = x^2+y^2\,,
$$
first proving that such system is superintegrable and then solving the
equations of motion using the Lagrangian formalism as an approach;
the third part deals with the Hamiltonian formalism and the $\la$-dependent
Hamilton-Jacobi equation which is shown to be separable in three different
coordinate systems.
In Sec.\ 4 we study the $\la$-dependent $n$-dimensional nonlinear oscillator
described by the Lagrangian
$$
  L  =  \frac{1}{2}\,\Bigl(\frac{1}{1 + \la\,r^2} \Bigr)
  \Bigl[\,\sum_i v_i^2  + \la\,\sum_{\,i<j}  J_{ij}^2 \,\Bigr]
  - \frac{\al^2}{2}\Bigl(\frac{r^2}{1 + \la\,r^2} \Bigr)\,,\quad
  r^2 = \sum_i x_i^2 \,,
$$
and we solve the associated equations obtaining different types of
solutions depending of the values of $\la$.
Next we consider the Hamiltonian approach,
we prove that it is a super-integrable system and we obtain different families
of constants of motion; the existence of several different sets of 
$n$ commuting
integrals is also discussed.
In Sec.\ 5 we start with a discussion of Lie algebra structure of the
symmetries of the $n=2$ non-linear oscillator;
in fact we see that they span a three-dimensional real Lie algebra
isomorphic to $SO(3,{\IR})$, $SO(2,1)$ or the Euclidean group in two 
dimensions,
depending on the sign of the parameter $\lambda$.
Then we present a geometric approach which explains the surprising
properties this $\la$-dependent system has and relates it to the
harmonic oscillator in spaces of constant curvature studied in Ref.\
\cite{RaS02a}-\cite{RaS03}.
Finally, in Sec.\ 6 we make some final comments.

%-----------------------------------------------------------
%%%%(Section 2.)
\section{$\la$-dependent ``Free Particle"}

  We will make use of the Lagrangian formalism as an approach.
That is, we will look, in the first place, for a Lagrangian function
$L(\la)$ with appropriate properties and then, we will turn our attention
to the corresponding nonlinear equations arising from $L(\la)$.

  It is clear that the equation (1) represents a non-linear version of a
linear equation with a non-linearity introduced by the coefficient $\la$;
but the important point is that, in Lagrangian terms, this coefficient
$\la$ modifies not only the quadratic potential $V=(1/2)\,x^2$ of the harmonic
oscillator but also the kinetic term $T=(1/2)\,v_x^2$.
Therefore, this particular system is not directly related with Henon-Heiles
or any other similar non-linear system \cite{Sa91}-\cite{CaRa99}
where the nonlinearity is introduced by just adding a new term of
higher order to the original potential.
Our first aim is to extend to $n=2$ dimensions this system in such
a way that its distinguishing properties are maintained.
So in order to construct the appropriate two-dimensional Lagrangian,
we may split the problem in two: the problem of the  kinetic term and
the problem of the potential.

  Next we begin with that of the kinetic term.

  Before giving the expression for the two--dimensional kinetic term,
let us consider some properties of the one-dimensional free-particle
motion characterized by the following Lagrangian
\begin{equation}
   L(x,v_x;\la) = T_1(\la)  = \frac{1}{2}\,\Bigl(\frac{v_x^2}{1 + 
\la\,x^2} \Bigr)\,,
\end{equation}
and the following equation
\begin{equation}
  (1 +\la x^2)\,\ddot{x} - (\la x)\,\dot{x}^2  = 0 \,,\quad\la>0\,.
\label{eqn1T1}\end{equation}
Two important properties are:
(i) The function $T_1(\la)$ is invariant under the action of the vector field
$X_x=X_x(\la)$ given by
$$
  X_x(\la) = \sqrt{\,1+\la\,x^2\,}\,\,\fracpd{}{x}  \,,
$$
in the sense that we have
$$
  X_x^t(\la)\Bigl(T_1(\la)\Bigr)=0  \,,
$$
where $X_x^t(\la)$ denotes the natural lift to the phase space $\IR{\times}\IR$
(tangent bundle in differential geometric terms) of the vector field 
$X_x(\la)$,
$$
  X_x^t(\la) = \sqrt{\,1+\la\,x^2\,}\,\,\fracpd{}{x}
  +  \Bigl(\frac{\la\,x v_x}{\sqrt{1+\la\,x^2\,}}\Bigr)\fracpd{}{v_x} \,.
$$
(ii) The general solution of the equation of motion, that can be 
directly obtained from
the conservation of the energy $E$, is given by
\begin{equation}
  x  = \Bigl(\frac{1}{\sqrt{\la}}\Bigr) \sinh\Bigl(\sqrt{2\la\,E}\,(t 
+ \phi)\Bigr)\,.
\label{xn1T1}\end{equation}
It is clear that this expression satisfies correctly the linear limit 
for $\la=0$.

  The first problem of the transition from 1-d to 2-d configuration space is the
construction  of the new two-dimensional $T_2(\la)$.  Many different
$\la$-dependent functions will have the same $\la=0$ limit,  so we 
must require that
the new function $T_2(\la)$ must satisfy  certain properties.
On the one hand we think that a natural requirement for $T_2(\la)$ is 
to  satisfy the
two-dimensional  versions of the previous points (i) and (ii).
On the other hand the potential $V(\la)$ for the two-dimensional
oscillator (to be studied in the next section) must be a $\la$-dependent
central potential such that the angular momentum be preserved;
but, according to the Noether theorem, exact symmetries of the potential
lead to constants of motion only if they also are symmetries of the
``kinetic energy".  Hence, a necessary condition must be that $T_2(\la)$
be also preserved with the same symmetry.

  We will consider as the starting point for our approach the following
three requirements.
%---------------
\begin{enumerate}
\item{}  The kinetic term $T_2(\la)$ must be a quadratic function of the
velocities that will remain invariant under rotations in the $\IR^2$ plane.
This means that it must depend of the coordinates $x$, $y$, by means of
$r^2=x^2+y^2$,
and of the velocities $v_x$, $v_y$, by means of $v_x^2+v_y^2$,
$(x v_y-yv_x)^2$, $(x v_x+yv_y)^2$ and  $(x v_x+yv_y)(x v_y - y v_x)$.
\item{}  $T_2(\la)$ should be invariant under (the lifts
of) the two vector fields $X_1(\la)$ and $X_2(\la)$ given by
%---------------
\begin{eqnarray}
   X_1(\la) &=& \sqrt{\,1+\la\,r^2\,}\,\,\fracpd{}{x}  \,,\cr
   X_2(\la) &=& \sqrt{\,1+\la\,r^2\,}\,\,\fracpd{}{y}  \,,{\nonumber}
\end{eqnarray}
%---------------
that represent the natural extension to $\IR^2$ of the vector field $X_x(\la)$
associated to $T_1(\la)$ in the one-dimensional case.
Notice that the previous point (1) implies that if $T_2(\la)$ is
invariant under (the lift of) $X_1(\la)$ then so will be under (the 
lift of) $X_2(\la)$.
\item{}  It must lead to a solution $(x(t), y(t))$ in $\IR^2$ for the
two-dimensional free-particle mo\-tion similar to the hyperbolic function
$x(t)$ in $\IR$ given by (\ref{xn1T1}), for the one-dimensional free-particle
motion.
\end{enumerate}
%---------------

  The most direct and simplest generalization of the one-dimensional kinetic
term $T_1(\la)$ is given by
$$
  T(x,y,v_x,v_y;\la) = (\frac{1}{2})\,
\Bigl(\frac{v_x^2 + v_y^2}{1 + \la\,(x^2+y^2)}\Bigr)\,,
$$
but this function does not satisfy the point (2). So let us try a more general
expression given by
$$
  T(x,y,v_x,v_y;\la) = (\frac{1}{2})\,\frac{W(v_x,v_y)}{1 + \la\,r^2}  \,,
$$
where $W=W(v_x,v_y)$ denotes
$$
  W = c_1(v_x^2 + v_y^2) + c_2(x v_y - y v_x)^2 +
        c_3(x v_x+yv_y)^2  + c_4 (x v_x+yv_y)(x v_y - y v_x)  \,,
$$
then the point (2) is satisfied only if
$$
  c_2 = \la\,c_1\,,\quad c_3=0\,,\quad c_4=0.
$$
Hence, we will choose the following two-dimensional kinetic function
%---------------
\begin{equation}
  T_2(\la) = (\frac{1}{2})\,\Bigl(\frac{1}{1 + \la\,r^2}\Bigr)\,
  \Bigl[\,v_x^2 + v_y^2 + \la\,(x v_y - y v_x)^2 \,\Bigr] \,,\quad
  r^2 = x^2+y^2     \,,
\label{Tn2}\end{equation}
%---------------
as the appropriate one for the two-dimensional $\la$-dependent dynamics.
Notice that this means that the $\la$-dependence is introduced in two
different ways:
the first one is the global factor $1/(1 + \la\,r^2)$ that is the most
direct $n=2$ extension of the one-dimensional factor $1/(1 + \la\,x^2)$
in (\ref{Lagn1});
the other $\la$-term is not so simple and it represents a two-dimensional
contribution that was not present (in fact, it can not be defined) in the
one-dimensional case.
Although one could guess that this additional term can introduce
difficulties, we will see that it really simplifies most of properties,
mainly all those  related with symmetries
(the geometric aspects will be discussed in Sec.\ 5). We also point out that we
admit $\la$ can take both positive and negative values.   It is clear that for
$\la<0$,
$\la=-\,|\la|$, the function (and the associated  dynamics) will have 
a singularity
at $1 -\,|\la|\,r^2=0$; because of this  we will restrict the study 
of the dynamics
to the interior of the circle
$x^2+y^2<1/|\la|$ that is the region in which $T_2(\la)$ is positive definite.

  It is known that a symmetric bilinear form in the velocities $(v_x,v_y)$
can be considered as associated to a two-dimensional metric $ds^2$ in $\IR^2$.
In this particular case, the function $T_2(\la)$ considered as a bilinear form
determines the following $\la$-dependent metric
%---------------
\begin{equation}
  ds^2(\la) = \Bigl(\frac{1}{1 + \la\,r^2}\Bigr)\,
  \Bigl[\,(1 + \la\,y^2)\,dx^2 + (1 + \la\,x^2)\,dy^2 - 2 \la\,x y \,dx
\,dy\,\Bigr]\,.
\label{ds2}\end{equation}
%---------------
This relation between kinetic term and metric implies that the 
Killing vectors of the
$\la$-metric $ds^2(\la)$ coincide with the exact Noether symmetries
of the $\la$-dependent free motion, that is, of the dynamics determined by
assuming the kinetic term as Lagrangian, $L(\la) = T_2(\la)$.

$T_2(\la)$ remains invariant under the actions of the lifts of the vector
fields $X_1(\la)$, $X_2(\la)$, and $X_J$, given by
%---------------
\begin{eqnarray}
   X_1(\la) &=& \sqrt{\,1+\la\,r^2\,}\,\,\fracpd{}{x}  \,,\cr
   X_2(\la) &=& \sqrt{\,1+\la\,r^2\,}\,\,\fracpd{}{y}  \,,\cr
   X_J   &=& x\,\fracpd{}{y} - y\,\fracpd{}{x}\,,   {\nonumber}
\end{eqnarray}
%---------------
in the sense that, if we denote by $X_r^t$, $r=1,2,J$, the natural lift to the
tangent bundle (phase space $\IR^2{\times}\IR^2$) of the vector field $X_r$,
%---------------
\begin{eqnarray}
   X_1^t(\la) &=& \sqrt{\,1+\la\,r^2\,}\,\,\fracpd{}{x}
   +  \la\,\Bigl(\frac{x v_x + y 
v_y}{\sqrt{1+\la\,r^2\,}}\Bigr)\fracpd{}{v_x} \,,\cr
   X_2^t(\la) &=& \sqrt{\,1+\la\,r^2\,}\,\,\fracpd{}{y}
   +  \la\,\Bigl(\frac{x v_x + y 
v_y}{\sqrt{1+\la\,r^2\,}}\Bigr)\fracpd{}{v_y} \,,\cr
{\nonumber}\end{eqnarray}
%---------------
then the Lie derivatives of $T_2(\la)$ with respect to $X_r^t(\la)$
vanish, that is
$$
  X_r^t(\la)\Bigl(T_2(\la)\Bigr)=0  \,,\quad X_J^t\Bigl(T_2(\la)\Bigr)=0
  \,,\quad  r=1,2.
$$

We close this section by solving the two-dimensional free-particle
motion determined by two-dimensional kinetic function (\ref{Tn2}).

  The Euler equations arising from $L(\la) = T_2(\la)$, and
representing the $n=2$ generalization of (\ref{eqn1T1}), are
%---------------
\begin{eqnarray}
  &&(1 + \la\,r^2)\,\ddot{x}
   - \la\,\bigl[\,\dot{x}^2 + \dot{y}^2 + \la\,(x\dot{y} - 
y\dot{x})^2\,\bigr]\,x =
0\,,\cr
  &&(1 + \la\,r^2)\,\ddot{y}
   - \la\,\bigl[\,\dot{x}^2 + \dot{y}^2 + \la\,(x\dot{y} - y\dot{x})^2\,\bigr]\,y 
=
0\,.
\label{eqn2T2}\end{eqnarray}
Of course they are much more difficult of solving than the single
equation (\ref{eqn1T1}); but assuming hyperbolic/trigonometric expressions
for the two functions, $x(t)$ and $y(t)$, then we obtain that the
solutions of (\ref{eqn2T2}) are given by
%---------------
\begin{eqnarray}\openup6pt
  &&x = \Bigl(\frac{A}{\sqrt{\la}}\Bigr)\sinh\,(C t + \phi_1)  \,,\quad
  y   = \Bigl(\frac{B}{\sqrt{\la}}\Bigr)\sinh\,(C t + \phi_2) 
\,,\quad \la>0\,, \cr
  &&x = \Bigl(\frac{A}{\sqrt{|\la|}}\Bigr)\sin\,(C t + \phi_1) \,,\quad
  y   = \Bigl(\frac{B}{\sqrt{|\la|}}\Bigr)\sin\,(C t + \phi_2) \,,\quad \la<0\,,
\label{xyT2sinh}
\end{eqnarray}
%---------------
with the only restriction
%---------------
\begin{eqnarray}
&& A^2 + B^2 + A^2 B^2 \sinh^2(\phi_1-\phi_2) = 1\,,\quad \la>0\,, \cr
&& A^2 + B^2 - A^2 B^2 \sin^2(\phi_1-\phi_2)  = 1\,,\quad \la<0\,.
{\nonumber}\end{eqnarray}
After some calculus we arrive, for $\la>0$ to
$$
  \la\,P_1^2 = C^2 (1 - B^2) \,,\quad
  \la\,P_2^2 = C^2 (1 - A^2) \,,\quad
  C^2 = 2 \la\,E   \,,
$$
hence the solutions $(x(t),y(t))$ reduce to
%---------------
\begin{eqnarray}
  x  &=& \sqrt{\frac{P_1^2 - \la\,J^2}{2\la\,E}}\,
\sinh\Bigl(\sqrt{2\la\,E}\,(t + \phi_1)\Bigr) \,,\cr
  y  &=& \sqrt{\frac{P_2^2 - \la\,J^2}{2\la\,E}}\,
\sinh\Bigl(\sqrt{2\la\,E}\,(t + \phi_2)\Bigr)\,,
\label{xyT2sinhb}\end{eqnarray}
for $\la>0$, where $J$ denotes the angular momentum $x v_y - y v_x$.
For the case $\la<0$ similar reasoning leads to
%---------------
\begin{eqnarray}
  x  &=& \sqrt{\frac{P_1^2 + |\la|\,J^2}{2|\la|\,E}}\,
\sin\Bigl(\sqrt{2|\la|\,E}\,(t + \phi_1)\Bigr) \,,\cr
  y  &=& \sqrt{\frac{P_2^2 + |\la|\,J^2}{2|\la|\,E}}\,
\sin\Bigl(\sqrt{2|\la|\,E}\,(t + \phi_2)\Bigr)\,.
\label{xyT2sinb}\end{eqnarray}
These results generalize the hyperbolic solution $x(t)$ in $\IR$
given by  (\ref{xn1T1}) and satisfy correctly the linear limit for $\la=0$.

%-----------------------------------------------------------
%%%%(Section 3.)
\section{$\la$-dependent $n=2$ quasi-Harmonic Oscillator}

  In this section we will study and solve the appropriate  $n=2$
versions of the $\la$-dependent equation (\ref{eq1}) and the
$\la$-dependent Lagrangian (\ref{Lagn1}).

%%%% (3.1)
\subsection{Noether symmetries of $\la$-dependent potentials}

A general standard $\la$-dependent Lagrangian (kinetic term minus a potential)
will have the following form
$$
  L(x,y,v_x,v_y;\la) = T_2(\la) - V(x,y;\la)
$$
in such a way that for $\la=0$ we recover the non-deformed linear system.

  It is known that if a potential $V(x,y)$, defined the Euclidean plane,
is invariant under either translations or rotations then it
admits a Noether integral of first order in the velocities.
Now, in this $\la$-dependent case, we have also a rather similar situation,
but as the Lagrangian system is $\la$-dependent so are the transformations.
The infinitesimal transformations generated by $X_1(\la)$ are
$$
  x' = x + \ep\,\,{\de}x\,,{\quad} y'=y\,,{\qquad} 
{\de}x=\sqrt{\,1+\la\,r^2\,}\,,
$$
and can be interpreted as a $\la$-dependent version of the translations along
the $x$-axis.
Similarly the generator of the one-parameter group of $\la$-dependent
translations along the $y$-axis
$$
  x' = x\,,{\quad} y'=y + \ep\,\,{\de}y\,,{\qquad} 
{\de}y=\sqrt{\,1+\la\,r^2\,}\,,
$$
is the vector field $X_2(\la)$
(the generator of rotations remains $\la$-independent).
If we denote by $\te_L$ the Cartan semibasic one-form
%---------------
\begin{eqnarray}
  \te_L &=& \Bigl(\fracpd{L}{v_x}\Bigr)\,dx + 
\Bigl(\fracpd{L}{v_y}\Bigr)\,dy \cr
        &=& \Bigl(\frac{1}{1 + \la\,r^2}\Bigr)
  \Bigl[\,v_x\,dx + v_y\,dy + \la\,(x v_y - y v_x)(x\,dy-y\,dx) 
\,\Bigr] \nonumber\,,
\end{eqnarray}
then we have the following
%---------------
\begin{enumerate}
\item{}  If the potential $V(\la)$ does not depend on the variable $x$ then
the Lagrangian $L(\la)$ is invariant under the transformations generated
by $X_1(\la)$;
in this case the function $P_1(\la)$ given by
$$
  P_1(\la) = i\Bigl(X_1^t(\la)\Bigr)\,\theta_L
           = \frac{v_x - \la\,J y}{\sqrt{\,1 + \la\,r^2\,}}
$$
is a constant of motion. Notice that in this case the coordinate $x$ is not
cyclic since it is always present in kinetic term $T_2(\la)$.
\item{}  If the potential $V(\la)$ is independent of the variable $y$ then
the Lagrangian $L(\la)$ is invariant under the transformations generated
by $X_2(\la)$;
in this case the function $P_2(\la)$ given by
$$
  P_2(\la) = i\Bigl(X_2^t(\la)\Bigr)\,\theta_L
           = \frac{v_y + \la\,J x}{\sqrt{\,1 + \la\,r^2\,}}
$$
is a constant of motion.
This situation is similar to the previous one; that is, the coordinate $y$
is not in the potential but, for $\la{\ne}0$, it appears in the kinetic term.
\item{} If $V(\la)$ is a central potential, then
$$
  J = i(X_J^t)\,\theta_L = x v_y - y v_x
$$
is a constant of motion.
Notice that both the vector field $X_J$ and $J$ are $\la$-independent.
\end{enumerate}
%---------------
In these three very particular cases, the corresponding system becomes
integrable with a second integral, $P_1(\la)$, $P_2(\la)$, or $J$,
arising from an exact Noether symmetry.

%---------------
%%%% (3.2)
\subsection{Lagrangian approach}

Let us consider the following $\la$-dependent Lagrangian
\begin{equation}
  L  = T_2(\la) - V_2(r;\la) \,,\quad
  V_2(r;\la)  = \frac{\al^2}{2}\,\Bigl(\frac{r^2}{1 + \la\,r^2} \Bigr) \,,\quad
  r^2 = x^2+y^2     \,,
\label{LagT2V2}\end{equation}
where $V_2(r;\la)$ is the direct extension to $n=2$ of the $n=1$ potential in
(\ref{Lagn1}).
The dynamics is given by the following $\la$-dependent vector field
$$
  \Ga_{\la} = v_x\,\fracpd{}{x} + v_y\,\fracpd{}{y}
            + F_x(x,y,v_x,v_y;\la)\,\fracpd{}{v_x}
            + F_y(x,y,v_x,v_y;\la)\,\fracpd{}{v_y}
$$
where the two functions $F_x$ and $F_y$ are given by
%---------------
\begin{eqnarray}
  F_x &=& -\,\al^2\Bigl(\frac{x}{1 + \la\,r^2}\Bigr)
         + \la\,\bigl[\,v_x^2 + v_y^2 + 
\la\,J^2\,\bigr]\Bigl(\frac{x}{1 + \la\,r^2}\Bigr) \,,\cr
  F_y &=& -\,\al^2\Bigl(\frac{y}{1 + \la\,r^2}\Bigr)
         + \la\,\bigl[\,v_x^2 + v_y^2 + 
\la\,J^2\,\bigr]\Bigl(\frac{y}{1 + \la\,r^2}\Bigr) \,,
{\nonumber}\end{eqnarray}
in such a way that for $\la=0$ we recover the dynamics of the standard
$2$-d harmonic oscillator
$$
  \Ga_0 = v_x\,\fracpd{}{x} + v_y\,\fracpd{}{y}
            -\,(\al^2 x)\,\fracpd{}{v_x} -\,(\al^2 y)\,\fracpd{}{v_y} \,.
$$
We see that for $\la<0$ the potential $V_2(r;\la)$ is a well with a 
boundless wall
at $r^2={1/|\la|}$; therefore,  all the trajectories will be bounded.
For $\la>0$ we have that $V_2(r;\la)\,{\to}\,(1/2)(\al^2/\la)$ for 
$r\,{\to}\,\infty$;
so for small energies the trajectories will be bounded but for 
$E(\la)>(1/2)(\al^2/\la)$
the trajectories will be unbounded
(see Figures I and II; notice that we have plotted $V(x,\la)$ but the graph of
$V_2(r;\la)$ is just the same but with $r{\ge}0$).

  Our objective is to solve the $\la$-dependent equations arising from
(\ref{LagT2V2}) and prove that this two-dimensional motion is periodic
in the bounded case as it was in the one-dimensional case.
At this point we recall that a system is called super-integrable if
it is integrable (in the sense of Liouville-Arnold) and, in addition,
possesses more independent first integrals than degrees of freedom;
in particular, if a system with $n$ degrees of freedom possesses
$N=2n-1$ independent first integrals, then it is called maximally
super-integrable.
An important property is that the existence of periodic motions is a
characteristic related  with  super-integrability; thus we may suspect
that this system is super-integrable.
This is actually the case, and  the following proposition states the
super-integrability of this $\la$-deformed system and proves the existence
of a complex factorization.
%---------------
%%%%%%%{Proposicion 1}
\begin{proposicion}
Let $K_1$, $K_2$, be the following two functions
\begin{eqnarray}
   K_1  &=& P_1(\la) + {\ii}{\al}\,\Bigl(\frac{x}{\sqrt{\,1 + 
\la\,r^2\,}}\Bigr) \,,\cr
   K_2  &=& P_2(\la) + {\ii}{\al}\,\Bigl(\frac{y}{\sqrt{\,1 + \la\,r^2\,}}\Bigr) 
\,.
{\nonumber}\end{eqnarray}
Then the complex functions  $K_{ij}$ defined as
$$
  K_{ij} = K_i\,K_j^* \,,\quad i,j=1,2\,,
$$
are constants of motion.
\end{proposicion}
%---------------
{\it Proof:}
We begin our analysis by considering the action of the vector field $\Ga_\la$,
which represents the time-derivative,  on the two $\la$-dependent functions
$P_1(\la)$ and $P_2(\la)$.
They are given by
$$
  \frac{d}{d t}\,P_1(\la) = -\,\Bigl(\frac{\al^2}{1 + \la\,r^2}\Bigr)
  \Bigl(\frac{x}{\sqrt{\,1 + \la\,r^2\,}}\Bigr) \,,{\qquad}
  \frac{d}{d t}\,P_2(\la) = -\,\Bigl(\frac{\al^2}{1 + \la\,r^2}\Bigr)
  \Bigl(\frac{y}{\sqrt{\,1 + \la\,r^2\,}}\Bigr) \,.{\nonumber}
$$
In a similar way, the time-derivative of the two velocity-independent
functions,  $x/\sqrt{\,1 + \la\,r^2}$ and $y/\sqrt{\,1 + \la\,r^2}$,
is given by
$$
  \frac{d}{d t}\,\Bigl(\frac{x}{\sqrt{\,1 + \la\,r^2\,}}\Bigr)
  = \Bigl(\frac{1}{1 + \la\,r^2}\Bigr)\,P_1(\la)   \,,{\qquad}
  \frac{d}{d t}\,\Bigl(\frac{y}{\sqrt{\,1 + \la\,r^2\,}}\Bigr)
  = \Bigl(\frac{1}{1 + \la\,r^2}\Bigr)\,P_2(\la)   \,.{\nonumber}
$$
Thus, the time-evolution of the two functions, $K_1$ and $K_2$, becomes
\begin{eqnarray}
   \frac{d}{d t}\,K_1 \equiv \Ga_\la (K_1)
   &=& \frac{d}{d t}\,P_1(\la)
    +  {\ii}{\al} \frac{d}{d t}\,\Bigl(\frac{x}{\sqrt{\,1 + 
\la\,r^2\,}}\Bigr)  \cr
   &=& \Big(\frac{1}{1 + \la\,r^2}\Big)
       \Big({\ii}{\al} P_1(\la) -\, \frac{\al^2\,x}{\sqrt{\,1 + 
\la\,r^2\,}}\Big)
    =  \Big(\frac{{\ii}{\al}}{1 + \la\,r^2}\Big) K_1     \,,
{\nonumber}\end{eqnarray}
and a similar calculus leads to
$$
  \frac{d}{d t}\,K_2 \equiv \Ga_\la (K_2) =
  \Big(\frac{{\ii}{\al}}{1 + \la\,r^2}\Big) K_2 \,.
$$
Thus we obtain
$$
	 {d\over dt}\,(K_i\,K_j^*)  \equiv \Ga_\la (K_i\,K_j^*) =  0 
\,,\quad i,j=1,2\,,
$$
Therefore the potential $V_2(\la)$ is super-integrable with the following three
integrals of motion
$$
  I_1(\la) = |\,K_1\,|^2 \,,{\quad}
  I_2(\la) = |\,K_2\,|^2 \,,{\quad}
  I_3 = \Im(K_{12}) = {\al}\,(x v_y - y v_x)  \,.
$$
That is, the existence of an invariant second order tensor $K_{ij}$,
admitting a complex factorization \cite{Pe90,CaMR02}, is preserved by the
nonlinearity introduced by $\la$.

  The system of equations
%---------------
\begin{eqnarray}
  &&(1 + \la\,r^2)\,\ddot{x}
   - \la\,\bigl[\,\dot{x}^2 + \dot{y}^2 + \la\,(x \dot{y} - y 
\dot{x})^2\,\bigr]\,x + \al^2\,x = 0\,,\cr
  &&(1 + \la\,r^2)\,\ddot{y}
   - \la\,\bigl[\,\dot{x}^2 + \dot{y}^2 + \la\,(x \dot{y} - y 
\dot{x})^2\,\bigr]\,y + \al^2\,y = 0\,,
\label{eqT2V2}\end{eqnarray}
cannot be directly solved in a simple way as it was the 
one-dimensional equation.
Nevertheless we can solve these equations by assuming certain 
particular expressions
(with some undetermined coefficients) for the two functions $x(t)$ and $y(t)$.

{\bf (i) Bounded motions}:
Let us look for solutions with the following periodic form
\begin{equation}
  x  = A \sin(\om t + \phi_1) \,,\quad
  y  = B \sin(\om t + \phi_2) \,,
\label{xyseno}\end{equation}
where $A$, $B$, $\phi_1$, $\phi_2$, and $\om$ are real parameters.
Then the equations (\ref{eqT2V2}) reduce to
$$
  A R \sin(\om t + \phi_1) = 0 \,,\quad
  B R \sin(\om t + \phi_2) = 0 \,,
$$
where $R$ is given by
$$
  R = \al^2 - \om^2 - \la\,(A^2+B^2)\,\om^2 - \la^2 A^2 
B^2\,\om^2\sin^2\phi_{12}
   \,,\quad  \phi_{12} = \phi_1-\phi_2\,.
$$
Therefore the functions (\ref{xyseno}) are in fact solutions of (\ref{eqT2V2})
but with $\om$, that represents the angular frequency of the motion,
$\la$-related with the coefficient $\al$ of the potential
(that represents the frequency of the $\la=0$ linear oscillator) by
$$
  \al^2 = M\,\om^2 \,,\quad
  M = 1 + \la\,P_e \,,\quad P_e = A^2+B^2 + \la\,\bigl(A^2 
B^2\sin^2\phi_{12}\bigr) \,.
$$
Notice that the coefficient $M$ is positive even for $\la=-|\la|<0$ 
since in that
case the amplitudes $A$ and $B$ must satisfy $A^2+B^2<1/|\la|$.

Once we know the solution of the dynamics, we can obtain the constants values
of the three integrals of motion, $I_1$, $I_2$, an $J$; they are given by
%---------------
\begin{eqnarray}
  I_1 &=&  (1 + \la\,B^2\,\sin^2\phi_{12})\,A^2 \om^2 \,,\cr
  I_2 &=&  (1 + \la\,A^2\,\sin^2\phi_{12})\,B^2 \om^2 \,,\cr
   J  &=& -\,\om\,A B\,\sin\phi_{12}  \,.
{\nonumber}\end{eqnarray}
Using these expressions, we can obtain the values of the amplitudes $A$, $B$,
as functions of the integrals of motion; we have
$$
  A = \bigl(\frac{1}{\om}\bigr) \sqrt{I_1 - \la\,J^2}  \,,\quad
  B = \bigl(\frac{1}{\om}\bigr) \sqrt{I_2 - \la\,J^2}  \,,
$$
so that the total energy becomes
%---------------
\begin{eqnarray}
  E(\la)  &=& (\frac{1}{2})\,\Bigl( A^2+B^2 \Bigr)\,\om^2
  + (\frac{\la}{2})\,\Bigl(A B \sin\phi_{12}\Bigr)^2\,\om^2    \,,\cr
   &=& \Bigl(\frac{\al^2}{2}\Bigr)\,\Bigl(\frac{P_e}{1+\la\,P_e} \Bigr)
<  \frac{\al^2}{2\,\la} \,.
{\nonumber}\end{eqnarray}
%---------------

  Let us summarize. The four coefficients ($A$, $B$, $\phi_1$, $\phi_2$)
remain arbitrary, the trajectories are sine-like periodic motions having
the same frequency but (possibly) differing in amplitude and in phase,
and the trajectories are ``ellipses" in $(x,y)$ plane.
The situation is very similar to the one of the linear oscillator,
the main difference laying in the frequency $\om$ that is given by
$\om^2 = \al^2/M$ and it depends, therefore, on the position (initial data).
We have two possibilities:
%---------------
\begin{itemize}
\item{} If the parameter $\la$ is negative $\la<0$, then $\om>\al$.
\item{} If the parameter $\la$ is positive $\la>0$, then $\om<\al$.
\end{itemize}
%---------------
The energy can take any value for $\la<0$, and it is always bounded by
$E_{\al,\la}=(1/2)(\al^2/\la)$ for $\la>0$, with the value of
$\om$ going down when the energy $E(\la)$ approaches to this upper value.

{\bf (ii) Unbounded motions}: Let us analyze the solutions 
corresponding to $\la>0$,
and $E>E_{\al,\la}$.

If we assume the following expressions
\begin{equation}
  x  = A \sinh(\Om t + \phi_1) \,,\quad
  y  = B \sinh(\Om t + \phi_2) \,,
\label{xysenoh}\end{equation}
for $x(t)$ and $y(t)$, then we obtain that they are solutions of (\ref{eqT2V2})
with the condition that $\al$ and $\Om$ must be $\la$-related by
$$
  \al^2 = M\,\Om^2 \,,\quad
  M = - 1 + \la\,P_h \,,\quad P_h = A^2+B^2 + \la\,\bigl(A^2 
B^2\sinh^2\phi_{12}\bigr) \,.
$$
The constant values of the three functions, $I_1$, $I_2$, an $J$, are given by
%---------------
\begin{eqnarray}
  I_1 &=&  (1 + \la\, B^2\,\sinh^2\phi_{12})\,A^2 \Om^2 \,,\cr
  I_2 &=&  (1 + \la\,A^2 \,\sinh^2\phi_{12})\,B^2 \Om^2 \,,\cr
   J  &=& -\,\Om\,A B\,\sinh\phi_{12}  \,,
{\nonumber}\end{eqnarray}
and the coefficients $A$, $B$, can be rewritten as follows
$$
  A = (\frac{1}{\Om}) \sqrt{I_1 - \la\,J^2}  \,,\quad
  B = (\frac{1}{\Om}) \sqrt{I_2 - \la\,J^2}  \,,
$$
Finally, the total energy becomes
%---------------
\begin{eqnarray}
  E(\la)  &=& (\frac{1}{2})\,\Bigl( A^2+B^2 \Bigr)\,\Om^2
  + (\frac{\la}{2})\,\Bigl(A B \sinh\phi_{12}\Bigr)^2\,\Om^2    \cr
  &=& \Bigl(\frac{\al^2}{2}\Bigr)\,\Bigl(\frac{P_h}{\la\,P_h-\,1} \Bigr)
  > \frac{\al^2}{2\,\la} \,.
{\nonumber}\end{eqnarray}
Figure III shows the form of the potential for $\la>0$ and $\la<0$.

{\bf (iii) Limiting unbounded motions}:
Let us analyze the very particular case characterized by $\la>0$,
and $E=E_{\al,\la}$ with $E_{\al,\la}=(1/2)(\al^2/\la)$.

If we assume the following expressions
\begin{equation}
  x  = A_1 t + B_1 \,,\quad
  y  = A_2 t + B_2 \,,
\label{xyatb}\end{equation}
for the solutions, then we obtain that they are solutions of 
(\ref{eqT2V2}) with
the following $\la$-dependent restriction for the four coefficients 
$A_1$, $A_2$,
$B_1$, and $B_2$,
$$
  \al^2 = \la\,P_L \,,\quad P_L = A_1^2+A_2^2 + \la\,\bigl(A_2 B_1 - 
A_1 B_2 \bigr)^2 \,.
$$
In this particular case the three functions, $I_1$, $I_2$, an $J$, 
take the form
%---------------
\begin{eqnarray}
  I_1 &=&  A_1^2 + \la\,\bigl(A_2 B_1 - A_1 B_2 \bigr)^2 \,,\cr
  I_2 &=&  A_2^2 + \la\,\bigl(A_2 B_1 - A_1 B_2 \bigr)^2 \,,\cr
   J  &=&  A_2 B_1 - A_1 B_2  \,,
{\nonumber}\end{eqnarray}
and the coefficients $A_1$, $A_2$, can be rewritten as follows
$$
  A_1 = \sqrt{I_1 - \la\,J^2}  \,,\quad
  A_2 = \sqrt{I_2 - \la\,J^2}  \,.
$$
Finally, the total energy becomes
$$
  E(\la)  = (\frac{1}{2})\,\Bigl( A_1^2+A_2^2 \Bigr)
  + (\frac{\la}{2})\,\bigl(A_2 B_1 - A_1 B_2 \bigr)^2  = 
\frac{\al^2}{2\,\la} \,.
$$

  Thus, the linear functions (\ref{xyatb}) appear, in this nonlinear
system, as a border-line solution making  separation between two different
situations in the $\la>0$ case:
trigonometric periodic oscillations (\ref{xyseno}) for small energies
and hyperbolic unbounded (scattering) evolutions (\ref{xysenoh}) for
high energies.

%---------------
%%%% (3.3)
\subsection{Hamiltonian approach}

The Legendre transformation is given by
$$
  p_x = \frac{(1 + \la\,y^2) v_x - \la\,x y v_y}{1 + \la\,r^2} \,,\quad
  p_y = \frac{(1 + \la\,x^2) v_y - \la\,x y v_x}{1 + \la\,r^2} \,,
$$
so that the form of the angular momentum is preserved by the Legendre map,
in the sense that we have $x p_y - y p_x=x v_y - y v_x$  (notice that this fact
is consequence of the introduction of the term $J$ in the definition 
of  $T_2(\la)$),
and the general expression for a $\la$-dependent Hamiltonian becomes
%---------------
\begin{equation}
  H(x,y,p_x,p_y;\la) = \bigl(\frac{1}{2}\bigr)\,\Bigl[\,p_x^2 + p_y^2
  + \la\,(x p_x + y p_y)^2 \Bigr]
  + \bigl(\frac{1}{2}\bigr)\,{\al^2}\,V(x,y)
\label{HamT2V}\end{equation}
%---------------
and hence the associated Hamilton-Jacobi equation takes the form
$$
   \Bigl(\fracpd{S}{x}\Bigr)^2 + \Bigl(\fracpd{S}{y}\Bigr)^2
  + \la\,\Bigl(x\,\fracpd{S}{x} + y\,\fracpd{S}{y}\Bigr)^2 + \al^2 
V(x,y) = 2 E \,.
$$
This equation is not separable in $(x,y)$ coordinates
because of the $\la$-dependent term; indeed the $(x,y)$ coordinates 
are not even
orthogonal.  Nevertheless we will see that there exist three particular
orthogonal coordinate systems,  and three particular families of associated
potentials, in which the  Hamiltonian (\ref{HamT2V}) admits Hamilton-Jacobi
separability.  To find them, we look for two one-parameter families of curves
$f_1(x,y,c_1)=0$ and
$f_2(x,y,c_2)=0$,
$g(\la)$-orthogonal to the level set curves of $x$ and $y$, we obtain 
respectively,
$$
  x = c_1 \sqrt{\,1 + \la\,y^2\,} \,,{\qquad}{\rm and}{\qquad}
  y = c_2 \sqrt{\,1 + \la\,x^2\,} \,.
$$
These two expressions suggest us to consider two particular
systems,  that we will denote by $(z_x,y)$ and $(x,z_y)$, that can be seen
as $\la$-dependent deformations of the $(x,y)$
coordinates; the third system is just the polar coordinate system.

(i) In terms of the new coordinates
$$
(z_x,y) \,,\quad z_x = \frac{x}{\sqrt{\,1 + \la\,y^2\,}} \,,
$$
the Hamilton-Jacobi equation becomes
$$
    (1 + \la\,z_x^2)\Bigl(\fracpd{S}{z_x}\Bigr)^2
  + (1 + \la\,y^2)^2\Bigl(\fracpd{S}{y}\Bigr)^2 + \al^2\,(1 + \la\,y^2)V
  = 2 (1 + \la\,y^2) E
$$
so if the potential $V(x,y)$ can be written on the form
\begin{equation}
   V = \frac{W_1(z_x)}{\,1 + \la\,y^2} + W_2(y)
\end{equation}
then the equation becomes separable.
The potential is therefore integrable with the following two quadratic
integrals of motion
%---------------
\begin{eqnarray}
  I_1(\la) &=& (1 + \la\,r^2)p_x^2 + {\al^2} W_1(z_x) \,,\cr
  I_2(\la) &=& (1 + \la\,r^2)p_y^2 - \la\,J^2 +
  {\al^2} \Bigr(W_2(y) - \frac{\la\,y^2}{\,1 + \la\,y^2}\,W_1(z_x)\Bigl).
{\nonumber}\end{eqnarray}
%---------------

(ii) In terms of
$$
(x,z_y) \,,\quad z_x = \frac{y}{\sqrt{\,1 + \la\,x^2\,}} \,,
$$
which is the symmetric one of the previous change (i), the 
Hamilton-Jacobi equation
becomes
$$
    (1 + \la\,x^2)^2\Bigl(\fracpd{S}{x}\Bigr)^2
  + (1 + \la\,z_y^2)\Bigl(\fracpd{S}{z_y}\Bigr)^2 + \al^2\,(1 + \la\,x^2)V
  = 2 (1 + \la\,x^2) E,
$$
and, therefore, if the potential $V(x,y)$ can be written on the form
\begin{equation}
   V = W_1(x) + \frac{W_2(z_y)}{\,1 + \la\,x^2}
\end{equation}
then the equation becomes separable.
The potential is therefore integrable with the following two quadratic
integrals of motion
%---------------
\begin{eqnarray}
  I_1(\la) &=& (1 + \la\,r^2)p_x^2 - \la\,J^2 +
  {\al^2} \Bigr(W_1(x) - \frac{\la\,x^2}{\,1 + \la\,x^2}\,W_1(z_y)\Bigl) \,,\cr
  I_2(\la) &=& (1 + \la\,r^2)p_y^2 + {\al^2} W_2(z_y) .
{\nonumber}\end{eqnarray}
%---------------

(iii) Finally we may use polar coordinates $(r,\phi)$.
Here the $\la$-dependent Hamiltonian (\ref{HamT2V}) is given by
$$
  H(r,\phi,p_r,p_\phi;\la) = (\frac{1}{2})\,\Bigl[\,(1 + \la\,r^2)p_r^2 +
\frac{p_\phi^2}{r^2}
\Bigr]
  + (\frac{\al^2}{2})\,V(r,\phi)
$$
so that the Hamilton-Jacobi equation is given by
$$
    (1 + \la\,r^2)\Bigl(\fracpd{S}{r}\Bigr)^2
  + \frac{1}{r^2} \Bigl(\fracpd{S}{\phi}\Bigr)^2 + \al^2\,V(r,\phi) = 2 E  \,.
$$
Let us suppose that the potential $V$ takes the form
\begin{equation}
   V = F(r) + \frac{G(\phi)}{r^2}\,,
\end{equation}
then the equation admits separability
$$
  \Bigl[\,r^2(1 + \la\,r^2)\Bigl(\fracpd{S}{r}\Bigr)^2
  + r^2 \bigl(\,\al^2 F(r) - 2 E\bigr) \,\Bigr]
  + \Bigl[\,\Bigl(\fracpd{S}{\phi}\Bigr)^2  + \al^2\,G(\phi) \,\Bigr] = 0 \,.
$$
The potential $V$ is integrable with the following two quadratic integrals
of motion
%---------------
\begin{eqnarray}
  I_1(\la)  &=&  (1 + \la\,r^2)\,p_r^2 + \Bigl(\frac{1-r^2}{r^2}\Bigr)\,p_\phi^2
  + \al^2 \,\Bigl[\, F(r) + \Bigl(\frac{1-r^2}{r^2}\Bigr)\,G(\phi) 
\,\Bigr] \,,\cr
  I_2(\la)  &=&  p_\phi^2 + \al^2 G(\phi)      \,.
{\nonumber}\end{eqnarray}
%---------------
It is clear that f $G=0$ then $V$ is a $\la$-dependent central potential
and the function $I_2$ just becomes $I_2=p_\phi^2$.

In these three separable cases the potential $V(\la)$ is integrable with two
quadratic constants of motion, $I_1(\la)$ and $I_2(\la)$, in such a way that
$$
  H(\la) = \bigl(\frac{1}{2}\bigr) \Bigl(I_1(\la) + I_2(\la)\Bigr)
$$
That is, there exist three different ways in which the Hamiltonian $H(\la)$
admits a decomposition as sum of two integrals; notice that in the linear case
$\la=0$ they reduce to only two since the two first cases, (i) and 
(ii), coincide.

We recall that a potential $V$ is called super-separable if it is separable in
more than one system of coordinates \cite{FrMS65}-\cite{TeTW01}.
The potential
$$
  V_2(\la)  = \frac{\al^2}{2}\, \Bigl(\frac{x^2+y^2}{1 + \la\,(x^2+y^2)} \Bigr)
$$
can be alternatively written as follows
%---------------
\begin{eqnarray}
  V_2(\la)  &=& \frac{\al^2}{2}\,\Bigl(\frac{1}{1+\la\,y^2}\Bigr)
  \Bigl[\frac{z_x^2}{1+\la\,z_x^2}  +  y^2 \Bigr]      \cr
    &=&  \frac{\al^2}{2}\,\Bigl(\frac{1}{1+\la\,x^2}\Bigr)
  \Bigl[ x^2  +  \frac{z_y^2}{1+\la\,z_y^2} \Bigr]    \cr
    &=&  \frac{\al^2}{2}\,\Bigl(\frac{r^2}{1+\la\,r^2}\Bigr)\,.{\nonumber}
\end{eqnarray}
%---------------
Therefore, it is super-separable since it is separable in three different
systems of coordinates $(z_x,y)$, $(x,z_y)$, and $(r,\phi)$.
Because of this the Hamiltonian
%---------------
\begin{equation}
  H = \bigl(\frac{1}{2}\bigr)\,\Bigl[\,p_x^2 + p_y^2 + \la\,(x p_x + y 
p_y)^2 \Bigr]
  + \frac{\al^2}{2}\, \Bigl(\frac{x^2+y^2}{1 + \la\,(x^2+y^2)} \Bigr)
\label{HamT2V2}\end{equation}
%---------------
admits the following decomposition
$$
   H = H_1 + H_2 - \la H_3
$$
where the three partial functions $H_1$, $H_2$, and $H_3$, are given by
%---------------
\begin{eqnarray}
  H_1  &=& \frac{1}{2}\,\Big[(1 + \la\,r^2)\,p_x^2 + \al^2
  \Bigl(\frac{x^2}{1+\la\,r^2}\Bigr)  \Bigr]    \,,\cr
  H_2  &=&  \frac{1}{2}\,\Big[(1 + \la\,r^2)\,p_y^2 + \al^2
  \Bigl(\frac{y^2}{1+\la\,r^2}\Bigr)  \Bigr]    \,,\cr
  H_3  &=&  \frac{1}{2}\,\bigl(x p_y - y p_x\bigr)^2 \,,{\nonumber}
\end{eqnarray}
%---------------
each one of these three terms has a vanishing Poisson bracket with $H$
for any value of the parameter $\la$
$$
  \bigl\{H\,,H_1 \bigr\} = 0\,,{\hskip 10pt}
  \bigl\{H\,,H_2 \bigr\} = 0\,,{\hskip 10pt}
  \bigl\{H\,,H_3 \bigr\} = 0\,.
$$
So the total Hamiltonian can be written as a sum, not of two,
but of three integrals of motion.
The third one, which represents the contribution of the angular momentum
$J$ to $H$, has the parameter $\la$ as coefficient; so it vanishes
in the linear limit $\la\to 0$.

%-----------------------------------------------------------
%%%%(Section 4.)
\section{$n$-dimensional Oscillator}

We have studied with a great detail the $\la$-dependent nonlinear
bi-dimensional oscillator; nevertheless, it is clear that this particular
Lagrangian (Hamiltonian) system admits a direct extension to $n$ dimensions.

%---------------
%%%% (4.1)
\subsection{Lagrangian formalism}

The $n$-dimensional $\la$-dependent Lagrangian is given by
%---------------
\begin{equation}
  L(\la)  =  \frac{1}{2}\,\Bigl(\frac{1}{1 + \la\,r^2} \Bigr)\,
  \Bigl[\,\sum_i v_i^2  + \la\,\sum_{\,i<j}  J_{ij}^2 \,\Bigr]
    - \frac{\al^2}{2}\,\Bigl(\frac{r^2}{1 + \la\,r^2} \Bigr)
\label{LagTnVn}\end{equation}
%---------------
where we have made use of the notation
$$
  r^2 = \sum_i x_i^2\,,\quad {\rm and}\quad J_{ij} =  x_i v_j - x_j v_i
  \,,\quad i,j = 1,\dots, n  \,.
$$
The $\la$-dependent Euler-Lagrange vector field $\Ga_{\la}$ takes the form
$$
  \Ga_{\la} = v_k\,\fracpd{}{x_k}  + F_k(x,v,\la)\,\fracpd{}{v_k}
$$
where the $n$ functions $F_k=F_k(x,v,\la)$, $k=1,\dots,n$, are given by
$$
  F_k = -\,\al^2\Bigl(\frac{x_k}{1 + \la\,r^2}\Bigr)
      + \la\,\Bigl[\,\sum_i v_i^2 + \la\,\sum_{\,i<j} J_{ij}^2\,\Bigr]
  \Bigl(\frac{x_k}{1 + \la\,r^2}\Bigr)    \,.
$$

Same as in the $n=2$ case, we can obtain the explicit expressions for the
solutions of the dynamics. We also have, in this $n$-dimensional case,
a quasi-harmonic oscillatory motion for $\la<0$, and two different
qualitative behaviours with a border-case when $\la>0$.

  (i) Periodic motions: We can assume that the solutions $x_i=x_i(t)$
of the $n$ equations
$$
  \ddot{x}_i  =  F_i(x,\dot{x};\la) \,,\qquad i = 1,\dots, n,
$$
are periodic functions of the form
$$
  x_i  =  A_i\,\sin(\om\,t + \phi_i)  \,,\qquad i = 1,\dots, n,
$$
Then we arrive at
$$
  A_i\,R\,\sin(\om\,t + \phi_i) = 0  \,,\qquad i = 1,\dots, n,
$$
where $R$ is given by
$$
  R = \al^2 - \Bigl(\,1 + \la\,\sum_i A_i^2\,\Bigr)\,\om^2
  - \la^2\,\Bigl(\,\sum_{i,j} A_i^2 A_j^2\sin^2\phi_{ij}\,\Bigr)\,\om^2  \,.
$$
Therefore, we have $\om^2 \ne \al^2$ but $\om^2 = \al^2/M$ with $M=M(\la)$
given by
$$
  M(\la) = 1 + \la\,P_e^{(n)} \,,\quad
  P_e^{(n)} = \sum_i A_i^2 + \la\,\Bigl(\,\sum_{i,j} 
A_i^2A_j^2\sin^2\phi_{ij}\,\Bigr)\,,
$$
and the energy $E(\la)$ is given by
%---------------
\begin{eqnarray}
  E(\la)  &=& (\frac{1}{2})\,\Bigl(\sum_i A_i^2 \Bigr)\,\om^2
  + (\frac{\la}{2})\,\sum_{i,j}\Bigl(A_iA_j \sin\phi_{ij}\Bigr)^2\,\om^2\,,\cr
   &=& 
\Bigl(\frac{\al^2}{2}\Bigr)\,\Bigl(\frac{P_e^{(n)}}{1+\la\,P_e^{(n)}} 
\Bigr) \,.
{\nonumber}\end{eqnarray}

  (ii) Unbounded motions:
If we assume that the solutions $x_i=x_i(t)$ are of the form
$$
  x_i  =  A_i\,\sinh(\Om\,t + \phi_i)  \,,{\qquad} i = 1,\dots, n,
$$
then we arrive to  $\Om^2 = \al^2/M$ with $M(\la)$ given by
$$
  M(\la) = - 1 + \la\,P_h^{(n)} \,,\quad
  P_h^{(n)} = \sum_i A_i^2 + \la\,\Bigl(\,\sum_{i,j} A_i^2 A_j^2 
\sinh^2\phi_{ij}\,\Bigr)\,,
$$
and the energy $E(\la)$ is given by
%---------------
\begin{eqnarray}
  E(\la)  &=& (\frac{1}{2})\,\Bigl( \sum_i A_i^2 \Bigr)\,\Om^2
  + (\frac{\la}{2})\,\sum_{i,j}\Bigl(A_iA_j \sin\phi_{ij}\Bigr)^2\,\Om^2    \cr
  &=& 
\Bigl(\frac{\al^2}{2}\Bigr)\,\Bigl(\frac{P_h^{(n)}}{\la\,P_h^{(n)}-\,1} 
\Bigr) \,.
{\nonumber}\end{eqnarray}

  (iii) Finally, in the positive $\la>0$ case, there exists a very particular
border-case that makes separation between the behaviours (i) and (ii).
It is characterized by the a value of the energy $E$ given by
$E=E_{\al,\la}$, and its time-evolution is represented by linear functions
$$
  x_i  = A_i t + B_i \,,{\qquad} i = 1,\dots, n,
$$
with the following $\la$-dependent restriction for the $2n$ 
coefficients $A_i$, $B_i$,
$$
  \al^2 = \la\,P_L \,,\quad
  P_L^{(n)} = \sum_i A_i^2 + \la\,\sum_{i,j}\bigl(A_i B_j - A_j B_i\bigr)^2 \,.
$$
Concerning the energy $E(\la)$ it takes  the particular value
$$
  E(\la)  = (\frac{1}{2})\,\sum_i A_i^2
  + (\frac{\la}{2})\,\sum_{i,j}\bigl(A_i B_j - A_j B_i\bigr)^2
  = \frac{\al^2}{2\,\la} \,.
$$

%---------------
%%%% (4.2)
\subsection{Hamiltonian formalism}

The $n$-dimensional $\la$-deformed Hamiltonian is given by
%---------------
\begin{equation}
  H  =  \frac{1}{2}\,\Bigl[\,\sum_i p_i^2 + \la\,(\sum_i x_i p_i)^2 \,\Bigr]
  + \frac{\al^2}{2}\,\Bigl(\frac{r^2}{1 + \la\,r^2} \Bigr)
  \,,{\quad}   r^2 = \sum_i x_i^2  \,.
\label{HamTnVn}\end{equation}
%---------------
It can be rewritten as a sum of $N= (1/2) n (n+1)$ quadratic terms as follows
$$
  H = \bigl(\frac{1}{2}\bigr) \Bigl( \sum_{k} I_k(\la) \Bigr)
  - \bigl(\frac{\la}{2}\bigr)\,\Bigl(\sum_{\,i<j} J_{ij}^2 \Bigr)
$$
where the functions $I_k(\la)$ and $J_{ij}$, which are given by
%---------------
\begin{eqnarray}
  I_k(\la) &=& (1 + \la\,r^2)\,p_k^2 + \al^2\Bigl(\frac{x_k^2}{1+\la\,r^2}\Bigr)
  \,,\quad   k=1,\dots,n, \cr
& \cr
  J_{ij} &=&  x_i p_j - x_j p_i \,,{\quad}  i,j=1,\dots,n,
\nonumber\end{eqnarray}
are constants of motion
$$
  \bigl\{H\,,I_k(\la)    \bigr\} = 0\,,{\hskip 20pt}
  \bigl\{H\,,J_{ij} \bigr\} = 0\,.
$$
Moreover, the $\la$-dependent functions
\begin{equation}
  I_{ij}(\la) =  (1 + \la\,r^2)\,p_i p_j + \al^2  \Bigl(\frac{x_i x_j}
  {1+\la\,r^2}\Bigr)
  \,,\quad i,j=1,\dots,n,\label{Integrals}
\end{equation}
are constants of motion as well.
Of course, the existence of these three different families,
$I_k(\la)$, $J_{ij}$, and $I_{ij}(\la)$, means that all these 
integrals cannot be
independent since the maximum number of
(time-independent) functionally  independent integrals is $N = 2 n - 1$.
In order to obtain a fundamental set of independent integrals
we can choose the $n$ functions $I_k(\la)$ and $n-1$ of the angular momenta;
an example is given by
$$
  (I_k(\la)\,,J_{i,i+1})\,,{\quad} k=1,\dots,n,{\quad} i=1,\dots,n-1.
$$
In fact this situation is rather the same that one finds in the linear
$\la=0$ case but with two important distinctions: first that all these
facts remain valid also for the unbounded (or scattering) motions, present when
$\la>0$ and  second that the algebra of Poisson brackets seems to be quadratic.

  We close this section observing that, if we call super-separable to a system
that admits Hamilton-Jacobi separation of variables (Schroedinger in the
quantum case) in more than one coordinate system, then quadratic
super-integrability (i.e., super-integrability with linear or
quadratic constants of motion) can be considered as a property
arising from super-separability.
In this $\la$-dependent case, the three families of constants of motion are
of such a class  ($J_{ij}$ are linear and $I_k(\la)$ and $I_{ij}(\la)$ are
quadratic), so we can conclude that the super-integrability of the
Hamiltonian (\ref{HamTnVn}) arises from its multiple separability.
In the linear $\la=0$ case, since the Hamiltonian is directly separable
in the $n$-dimensional cartesian system, the Hamiltonian is just the sum of
the $n$ partial one-dimensional energies; in the general nonlinear $\la{\ne}0$
case, the functions $I_i(\la)$, and $J_{ij}$ $i,j=1,\dots,n$, arise 
from separability
in the $n$-dimensional versions of the two-dimensional coordinates $(z_x,y)$
and $(x,z_y)$ studied in Sec. 3 and in the $n$-dimensional spherical system.
Finally, there exist many different sets of $n$ conmmuting constants;
as an example, in the $n=3$ case we have the following three sets of
involutive integrals
%---------------
\begin{eqnarray}
  &&(\,I_1\,,\,I_2-\la\,J_{12}^2\,,\,I_3-\la\,(J_{23}^2+J_{31}^2)\,) \,,\cr
  &&(\,I_1-\la\,(J_{12}^2+J_{31}^2)\,,\,I_2\,,\,I_3-\la\,J_{23}^2\,) \,,\cr
  &&(\,I_1-\la\,J_{31}^2\,,\,I_2-\la\,(J_{12}^2+J_{23}^2)\,,\,I_3\,)  \,,
\nonumber\end{eqnarray}
as well as $(I_1+I_2+I_3,J_{ij},J^2)$.
It is clear from this example that in the general $n$-dimensional case there
exist many more different ways of constructing involutive sets of $n$ 
integrals.

%-----------------------------------------------------------
%%%%(Section 5.)
\section{A Geometric Interpretation}

  In this section we will discuss some additional properties that will
prove to be related with a new geometric interpretation.
In particular, we will see that this quasi-harmonic nonlinear oscillator
turns out to be closely related with the harmonic oscillator on a space
of constant curvature.

%---------------
%%%% 5.1
\subsection{Some additional properties}

  We begin by considering two remarkable properties; the first one is
concerned with the symmetries of two-dimensional system and the second
one with the Lagrangian of the one-dimensional oscillator.

(i)
The three $\la$-dependent vector fields $X_1$, $X_2$, $X_J$, obtained
in Sec.\ 2 as the Killing vectors of the metric $ds^2$ or, equivalently,
as the Noether symmetries of the Lagrangian $L(\la) = T_2(\la)$ of the
$\la$-dependent free particle, close the following Lie algebra,
\begin{equation}
  [X_1(\la)\,,X_2(\la)] = {\la}\,X_J \,,{\quad}
  [X_1(\la)\,,X_J] =       X_2(\la)   \,,{\quad}
  [X_2(\la)\,,X_J] =    -\,X_1(\la)   \,.
\label{LieAlg}
\end{equation}
This means that the $2$-d configuration space has a three-dimensional
symmetry Lie algebra, hence a maximal one, and this implies that the space
should be of constant curvature.
Indeed the Lie algebra (\ref{LieAlg}) is  isomorphic to the Euclidean
algebra in the particular $\la=0$ case, and to the  Lie algebra of
the isometries of the two-dimensional spherical ($\la<0$) and hyperbolic
spaces ($\la>0$) in the general $\la{\ne}0$ case.

  It seems, therefore, that there exists a certain relation between
this $\la$-dependent nonlinear oscillator and the properties of the
two-dimensional spaces of constant curvature.
At this point we recall that we have proved in Sec.\ 4 that this
nonlinear system is well defined for any number $n$ of degrees of freedom;
therefore, it is natural to guess that if such relationship exists
then it must be true, not only for $n=2$, but for any dimension.

  Although the $n=1$ case can be considered as a very special case,
it seems convenient to go back to the one-dimensional oscillator and
analyze again its properties but now in relation with the above point (i).
We have obtained the following property that concerns the one-dimensional
nonlinear system.

(ii)
Let us consider the change $(x,v_x) \to (q,v_q)$  given by
$$
  q  = \Bigl(\frac{1}{\sqrt{\la}}\Bigr) \sinh^{-1}\bigl(\sqrt{\la}\,x\bigr)
  \,,{\quad} \la > 0 \,,
$$
then the Lagrangian
%---------------
\begin{equation}
  L(x,v_x;\la)  = \frac{1}{2}\,\Bigl(\frac{1}{1 + \la\,x^2} \Bigr)\,(v_x^2 -
\al^2\,x^2)
\label{LagRx}\end{equation}
%---------------
becomes
%---------------
\begin{equation}
  L(q,v_q;\la)  = \frac{1}{2}\,v_q^2 - \left(\frac{\al^2}{2\,\la}\right)
  \tanh^2(\sqrt{\la}\,q) \,.
\label{LagH1q}\end{equation}
%---------------
In the negative case, $\la<0$, $\la=-\,|\la|$, the corresponding change
is given by
$$
  q  = \Bigl(\frac{1}{\sqrt{|\la|}}\Bigr) \sin^{-1}\bigl(\sqrt{|\la|}\,x\bigr)
  \,,{\quad} \la=-\,|\la| \,,
$$
and then we arrive at
%---------------
\begin{equation}
  L(q,v_q;\la) = \frac{1}{2}\,v_q^2 - \left(\frac{\al^2}{2\,|\la|}\right)
   \tan^2(\sqrt{|\la|}\,q) \,.
\label{LagS1q}\end{equation}
%---------------
Hence, we can remove the $\la$ parameter from the kinetic energy $T_1(\la)$,
but the price for this simplification is that the potential $V$ drops
its rational character and becomes a trigonometric or hyperbolic squared
tangent function
%---------------
\begin{eqnarray}
  V(q;\la) &=&  \frac{1}{2}\Bigl(\frac{\al^2}{\la}\Bigr) \tanh^2(\sqrt{\la}\,q)
\,,{\quad}{\rm for}{\quad}  \la>0 \,,\cr
  V(q;\la) &=&  \frac{1}{2}\Bigl(\frac{\al^2}{|\la|}\Bigr) 
\tan^2(\sqrt{|\la|}\,q)
\,,{\quad}{\rm for}{\quad}  \la<0  \,.
{\nonumber}\end{eqnarray}
%---------------
 From a practical point of view, this change does not simplify really
the problem, since the two new equations
$$
  \sqrt{|\la|}\,\,\ddot{q} + \al^2\,F(q;\la)  = 0 \,,
$$
with $F=F(q,\la)$ given by
$$
  F =  \frac{\sinh\,(\sqrt{\la}\,q)}{\cosh^3(\sqrt{\la}\,q)}  \,,\quad  \la>0\,,
  {\quad}{\rm and}{\quad}
  F =  \frac{\sin\,(\sqrt{|\la|}\,q)}{\cos^3(\sqrt{|\la|}\,q)}\,,\quad  \la<0\,,
$$
are not easier of solving than the original one (\ref{eq1}).
Nevertheless, from a more qualitative viewpoint, the new aspect adopted by
this one-dimensional Lagrangian is a very interesting fact since the form
obtained for the new potential $V(q,\la)$ is closely related with the polar
coordinate $(\rho,\phi)$ expression for the two-dimensional
potential of the harmonic oscillator on two-dimensional spaces of constant
curvature, previously studied (by two of the present authors) in Refs.
\cite{RaS02a}-\cite{RaS03}.

  It is clear from properties (i) and (ii) that we can conjecture the
existence of a direct relationship between this particular nonlinear
oscillator and the harmonic oscillator on spaces of constant curvature.
Another remarkable property is that the set of integrals of motion
(\ref{Integrals})
is analogous to the ones appearing in the $n$-dimensional version of
the Smorodinsky-Winternitz system in curved spaces \cite{BHS03}.

%---------------
%%%% 5.2
\subsection{The harmonic oscillator on spaces of constant curvature}

  Higgs \cite{Hi79} and Leemon \cite{Le79} analyzed the
characteristics of the two fundamental central potentials,
Kepler problem and harmonic oscillator, on the $N$-dimensional sphere.
Since then a certain number of authors have studied this question
from both the classical and the quantum
points of view \cite{GrPS95b}-\cite{RSSBanach}.
Next we recall some of the basic properties of the formalism
studied in \cite{RaS02a}-\cite{RaS03}.

We begin for the following three basic ideas:
%---------------
\begin{itemize}
\item{} The harmonic oscillator is a system that is well defined in
all the three two-dimensional spaces of constant curvature
(sphere $S^2$, Euclidean plane $\IE^2$, and hyperbolic plane $H^2$).
\item{} A joint approach, where the usual Euclidean system and the two curved
systems  (defined on $S^2$ and $H^2$) can be studied, all the three, 
at the same
time, is possible using the curvature $\k$ as a parameter.
\item{} Thus the spherical and hyperbolic oscillators
can be considered as {\it curvature deformations} of the well known ``flat''
Euclidean  oscillator which arises as a very particular
case of the more general ``curved" systems.
\end{itemize}
%---------------
This two last points mean that all the theory must be developed by making
use of $\k$-dependent mathematical expressions leading to general
$\k$-dependent properties.
The specific properties characterizing the harmonic oscillator on the sphere,
on the Euclidean plane, or on the hyperbolic plane, are then obtained
particularizing for $\k>0$, $\k=0$, or $\k<0$.
  In order to present these expressions in a form which holds
simultaneously for any value of $\k$, the theory can be developed by
making use of the following ``tagged" trigonometric functions
%---------------
\begin{equation}
   \Cos_{\k}(x) = \cases{
   \cos{\sqrt{\k}\,x}       &if $\k>0$, \cr
   {\quad}  1               &if $\k=0$, \cr
   \cosh\!{\sqrt{-\k}\,x}   &if $\k<0$, \cr}{\qquad}
%---------------
   \Sin_{\k}(x) = \cases{
   \frac{1}{\sqrt{\k}} \sin{\sqrt{\k}\,x}     &if $\k>0$, \cr
   {\quad}   x                                &if $\k=0$, \cr
   \frac{1}{\sqrt{-\k}}\sinh\!{\sqrt{-\k}\,x} &if $\k<0$, \label{SkCk}\cr}
\end{equation}
%---------------

  If we make use of this $\k$-dependent notation in the polar coordinates
$(\rho,\phi)$ system, then the differential element of distance, on the three
spaces  $(S^2,\IE^2,H^2)$, can be written as follows
$$
  ds_{\k}^2 = d\rho^2 + \Sin_\k^2(\rho)\,d{\phi}^2 \,,
$$
It reduces to
$$
  ds_1^2 =    d\rho^2 + (\sin^2 \rho)\,d{\phi}^2 \,,{\quad}
  ds_0^2 =    d\rho^2 + \rho^2\,d{\phi}^2 \,,{\quad}
  ds_{-1}^2 = d\rho^2 + (\sinh^2 \rho)\,d{\phi}^2\,,
$$
in the three particular cases of the unit sphere, Euclidean plane,
and `unit` Lobachewski plane.
Consequently, the Lagrangian for the geodesic (free) motion
on the spaces $(S^2,\IE^2,H^2)$ is given by the kinetic term
arising from the metric
$$
  L(\k) = T(\k) = (\frac{1}{2})\,\left(v_\rho^2 + \Sin_\k^2(\rho) 
v_\phi^2\right)\,,
$$
and the Lagrangian for a general mechanical system (Riemmanian metric 
minus a potential)
has the following form
$$
  L(\k) = (\frac{1}{2})\,\left( v_\rho^2 + \Sin_\k^2(\rho) v_{\phi}^2 \right)
        -  U(\rho,\phi,\k)  \,.
$$
It is clear that the well known expression for a natural Lagrangian on
the Euclidean plane
$$
  L = (\frac{1}{2})\,(v_\rho^2 + \rho^2\,v_{\phi}^2) - V(\rho,\phi)
  \,,{\quad}   V(\rho,\phi) = U(\rho,\phi,0) \,,
$$
is obtained as the particular $\k=0$ case of $L(\k)$.

  Until now we have considered general aspects of the theory of
Lagrangian systems on curved spaces.
Now let us turn our attention to the spherical and hyperbolic
harmonic oscillators;
it is characterized by the following Lagrangian with curvature $\k$
$$
  L = (\frac{1}{2})\,\left(v_\rho^2 + \Sin_\k^2(\rho) v_\p^2\right)
    - (\frac{1}{2})\,\om_0^2\,\Tan_\k^2(\rho) .
$$
where the $\k$-dependent tangent $\Tan_\k(\rho)$ is defined in the natural way
$\Tan_\k(\rho) = \Sin_{\k}(\rho)/\Cos_{\k}(\rho)$.
In this way, the harmonic oscillator on the unit sphere (Higgs oscillator),
on the Euclidean plane, or on the unit Lobachewski plane, arise as 
the following
three particular cases
$$
  U_1(\rho) = (\frac{1}{2})\,\om_0^2\,\tan^2\rho     \,,{\quad}
  V(\rho) = U_0(\rho) = (\frac{1}{2})\,\om_0^2\,\rho^2   \,,{\quad}
  U_{-1}(\rho) = (\frac{1}{2})\,\om_0^2\,\tanh^2\rho \,.
$$
The Euclidean oscillator $V(\rho) = U_0(\rho)$ (parabolic potential without
singularities) appears in this formalism as making a separation 
between two different
situations.
The spherical potential is represented by a well with singularities
on the border (impenetrable walls at the equatorial circle
$\rho=\pi/2\sqrt{\k}$ if the potential center is placed  at the poles), and the
hyperbolic potential by a well with finite depth since for $\k<0$, $\k=-|\k|$,
we have $U_{\k}(\rho){\to}(1/2)(\om_0^2/|\k|)$ when $\rho\to\infty$.
Actually, the Scarf potential $V(x)=\gamma/\cos^2(x)$, which  differs 
in a constant
term from $U_1(x)$, has been studied in solid state physics and has many
interesting properties \cite{Sc58}-\cite{Gr93}.
Figure IV plots $U_{\k}(\rho)$, for the three particular cases of the 
unit sphere,
Euclidean plane, and `unit` Lobachewski plane;
notice the great resemblance with Figure III.

%---------------
%%%% 5.3
\subsection{On the existence of a relation between two dynamics}

  The abovementioned property (ii), for the $n=1$ case, suggests that the
Lagrangian (\ref{LagRx}) and the equation (\ref{eq1}) is a nonlinear
model for an harmonic oscillator in the circle $S^1$ ($\la<0$)
or in the hyperbolic line ($\la>0$).
In a similar way, if we consider the two properties (i) and (ii) together,
we arrive at the conclusion that this correspondence must also exist for
the nonlinear $n=2$ oscillator. Remark that in the $n=1$ case, $S^1$ and $H^1$
must be understood as one-dimensional spaces obtained by endowing each single
geodesic of $S^2$ or $H^2$ with the induced metric.
Motion on $S^1$ and $H^1$ will correspond to the $J=0$ radial motions on
$S^2$ or $H^2$.

  Hence, the Lagrangian (\ref{LagRx}) for the equation (\ref{eq1}) can be
considered as a nonlinear model on the $\IR$-line for the harmonic oscillator
on the circle $S^1$ and the hyperbolic line $H^1$;
the Lagrangian $L=T_2(\la)-V_2(\la)$ of equation (\ref{LagT2V2}) as a nonlinear
$\IR^2$--model for the harmonic  oscillator on the sphere $S^2$ and 
the hyperbolic
plane $H^2$;  finally, the more general $n$--dimensional Hamiltonian given by
equation (\ref{HamTnVn}) as a nonlinear $\IR^n$--model for the 
harmonic oscillator
on the $n$--dimensional  spaces $S^n$ and $H^n$.
So we arrive to the conclusion that an harmonic oscillator, that is a linear
system,  when is defined on a space of constant curvature turns out 
to be equivalent
to a nonlinear oscillator on $\IR$, $\IR^2$ or $\IR^n$, with the nonlinear
parameter $\la$ playing the role of the (negative of the) curvature $\k$.
Notice that, in dynamical terms, this equivalence is a relation of conjugacy,
and also that the existence of this relation can be considered as the origin
of the quasi-harmonic behaviour obtained for the solutions of the 
nonlinear system.

  The existence of this relationship makes this nonlinear $\IR^2$ (or $\IR^n$)
system even more interesting.
 From an abstract or qualitative viewpoint, the system on the sphere $S^2$
(or on the hyperbolic space $H^2$) may be considered as the more 
fundamental one;
nevertheless, we have been able to solve the equations and to obtain 
the explicit
solution for the dynamics because we were working with the nonlinear 
$\la$-dependent
$\IR^2$--model (or $\IR^n$--model).
Hence, the nonlinear system studied in this article results to be much
more appropriate for the explicit resolution of problems.

%-----------------------------------------------------------
%%%%(Section 6.)
\section{Final Comments and Outlook}

  We have proved that the Lagrangian (\ref{LagT2V2}) is the appropriate
two-dimensional version of the Lagrangian (\ref{Lagn1}), (\ref{LagRx}),
and we have solved the nonlinear equations (\ref{eqT2V2}) in the two cases
$\la<0$ and $\la>0$.
Moreover we have proved in Sec.\ 4 that this nonlinear system admits
a $n$--dimensional version given by (\ref{LagTnVn}) and (\ref{HamTnVn}),
and we have also solved the correspondent system of $n$ equations.

We think that all these results suggest the study of some related
questions among them we point out the following:
Firstly, the Poisson brackets of the integrals of motion for the
$2$--dimensional Hamiltonian (\ref{HamT2V2}) or the $n$--dimensional
Hamiltonian (\ref{HamTnVn}) seems to close a quadratic Poisson algebra
(see e.g. Ref.\ \cite{Dask01});
this is a very interesting problem that deserves to be studied.
Secondly, the geometric relations obtained in the last Sec.\ 5, is
a matter related with the existence of two different but
conjugate dynamical systems.
Conjugate systems are systems related by diffeomorphisms preserving
the fundamental properties; so it is convenient to develop a
deeper analysis of this particular relation of conjugacy existing
between the nonlinear $\la$-dependent oscillator and the harmonic
oscillator on the sphere $S^2$ or the hyperbolic plane $H^2$.
Thirdly, it is clear that the calculus are easier of handle when
working with the $\la$-dependent formalism than when working on
the constant curvature spaces;
therefore, when looking for new dynamical results on the spaces
such as $S^2$ or $H^2$, a practical strategy will be to first consider
the question by using Lagrangians such as
(\ref{LagT2V2}) or (\ref{LagTnVn}).
Fourthly, in Refs. \cite{RaS02b} and \cite{RaS03} we have studied
not only the central 1:1 oscillator given by the function
$V_\k(\rho)=(1/2)\,\om_0^2\,\Tan_\k^2(\rho)$
but also some others non-central non-isotropic oscillators
$V_\k(\rho,\phi)$;
it will be convenient to consider these non-central
problems also inside this $\la$-dependent formalism.
Finally, we mention the study of the quantized versions of all these
nonlinear systems.  Notice that, from the quantum point of view,
as it is an oscillator with a position-dependent effective mass,
there is a problem with the appropriate order for the factors in the
kinetic term.
We think that all these problems are examples of some
open questions that must be investigated.

{\small
\section*{\bf Acknowledgments.}
Support of projects  BFM-2003-02532, FPA-2003-02948,
BFM-2002-03773, and CO2-399 is acknowledged.
The work of M. Senthilvelan is supported by the Department of
Science and Technology, Government of India.

%-----------------------------------------------------------
 }
%-------------------------------------------------------------------
%---------------------------------------------------------------------
\vfill\eject
%---------------------------------------------------------------------
\section*{Figure Captions}

{\sc Figure I}.{\enskip}
Plot of $V(\la)=(1/2)\,(\al^2 x^2)/(1 + \la\,x^2)$, $\al=1$, $\la<0$,
as a function of $x$, for $\la=-2$
(upper curve), and $\la=-1$ (lower curve).

{\sc Figure II}.{\enskip}
Plot of $V(\la)=(1/2)\,(\al^2 x^2)/(1 + \la\,x^2)$, $\al=1$, $\la>0$,
as a function of $x$, for $\la=1$
(upper curve), and $\la=2$ (lower curve).

{\sc Figure III}.{\enskip}
Plot of $V_2(\la)=(1/2)\,(\al^2 r^2)/(1 + \la\,r^2)$, $\al=1$,
as a function of $r$, for $\la=-1$
(upper curve), $\la=0$ (dashed line) and $\la=1$ (lower curve).

{\sc Figure IV}.{\enskip}
Plot of $U_{\k}(\rho)=(1/2)\,\om_0^2\,\Tan_\k^2(\rho)$, $\om_0=1$, as 
a function of $\rho$,
for $\k=-1$ (lower curve), $\k=0$ (dash line), and $\k=1$ (upper curve).

%---------------------------------------------------------------------
\vfill\eject\null
%---------------------------------------------------------------------

\epsfbox{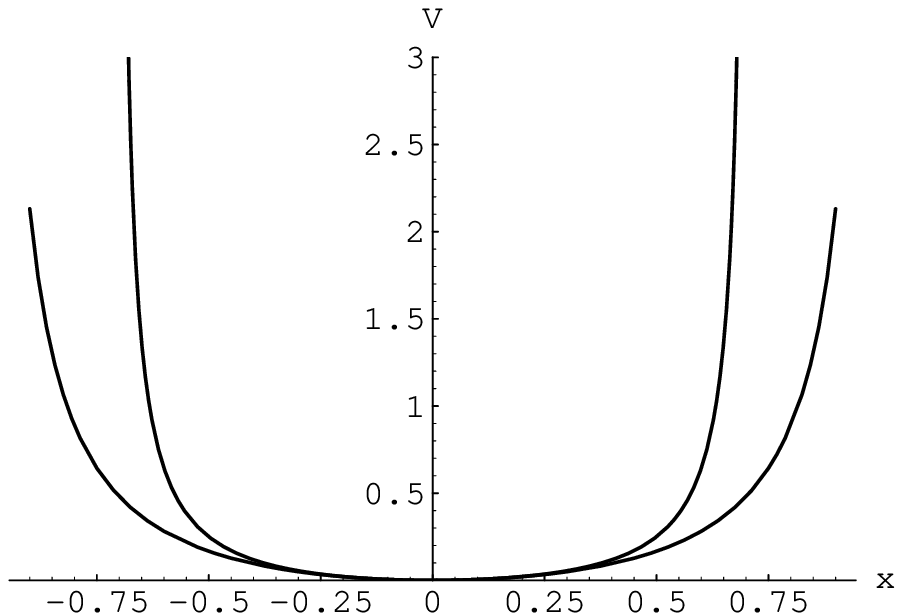}

{\medskip}
{\sc Figure I}.{\enskip}
Plot of $V(\la)=(1/2)\,(\al^2 x^2)/(1 + \la\,x^2)$, $\al=1$, $\la<0$,
as a function of $x$, for $\la=-2$
(upper curve), and $\la=-1$ (lower curve).

{\vskip 40pt}
\epsfbox{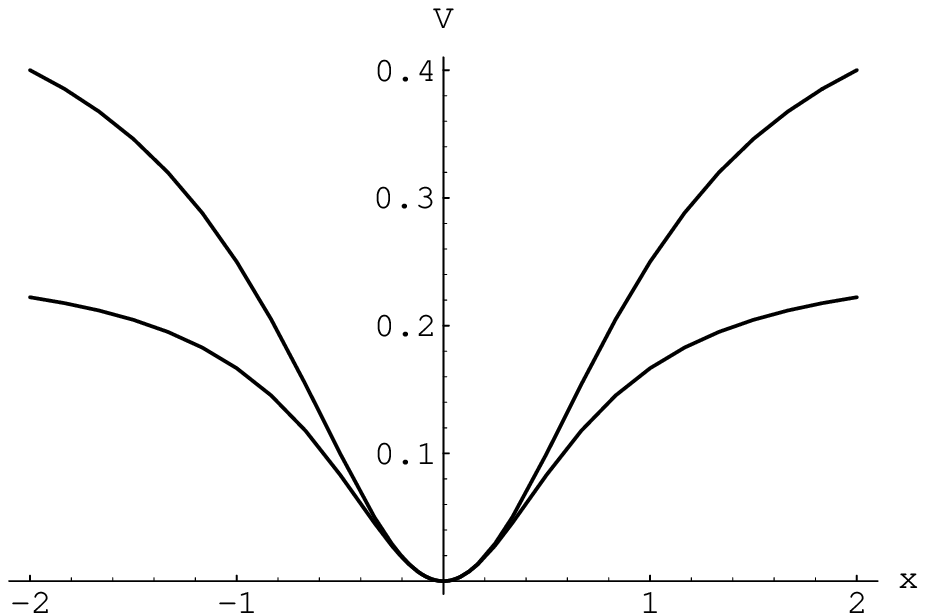}

{\medskip}
{\sc Figure II}.{\enskip}
Plot of $V(\la)=(1/2)\,(\al^2 x^2)/(1 + \la\,x^2)$, $\al=1$, $\la>0$,
as a function of $x$, for $\la=1$
(upper curve), and $\la=2$ (lower curve).

%--------
\vfill\eject
%--------

\epsfbox{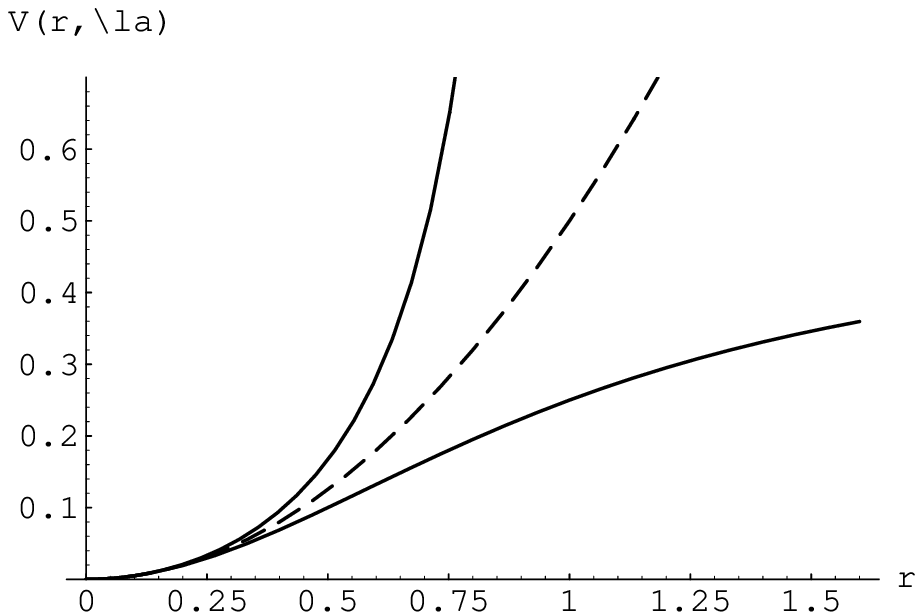}

{\medskip}
{\sc Figure III}.{\enskip}
Plot of $V_2(\la)=(1/2)\,(\al^2 r^2)/(1 + \la\,r^2)$, $\al=1$,
as a function of $r$, for $\la=-1$
(upper curve), $\la=0$ (dashed line) and $\la=1$ (lower curve).

{\vskip 40pt}
\epsfbox{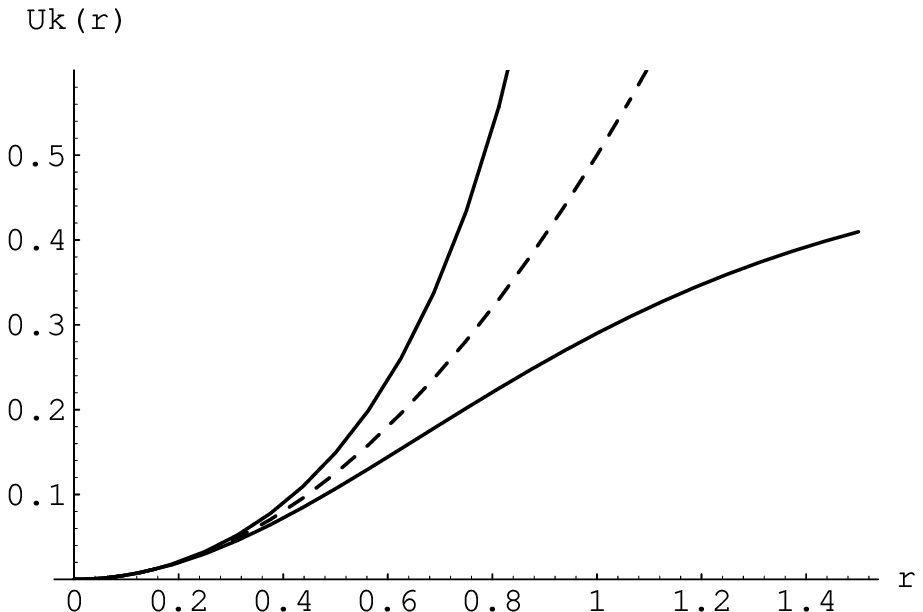}

{\medskip}
{\sc Figure IV}.{\enskip}
Plot of $U_{\k}(\rho)=(1/2)\,\om_0^2\,\Tan_\k^2(\rho)$, $\om_0=1$, as 
a function of $\rho$,
for $\k=-1$ (lower curve), $\k=0$ (dash line), and $\k=1$ (upper curve).

%---------------------------------------------------------------------
\vfill\eject
%---------------------------------------------------------------------

%-------------------------------------------------------------------

\begin{thebibliography}{99}
%-----------------------------------------------------------

%-----------------------------------------------------
\bibitem{MaL74}{P.M. Mathews and M. Lakshmanan},  1974
{\rm ``On a unique nonlinear oscillator"},
{\sl Quart. Appl. Math.} {\bf 32} 215--218.
%-----------------------------------------------------
\bibitem{LaRa03}{M. Lakshmanan and S. Rajasekar},  2003
{\sl ``Nonlinear dynamics. Integrability, Chaos and Patterns"},
Advanced Texts in Physics   (Springer-Verlag, Berlin).
%-----------------------------------------------------
\bibitem{DeSS69}{R. Delbourgo, A. Salam, and J. Strathdee}, 1969
{\rm ``Infinities of nonlinear and Lagrangian theories"},
{\sl Phys. Rev.} {\bf 187}, 1999--2007.
%-----------------------------------------------------
\bibitem{NiW72}{K. Nishijima and T. Watanabe},  1972
{\rm ``Green's functions in non-linear field theories"},
{\sl Prog. Theor. Phys.} {\bf 47}, 996--1003.
%--------------------------------------------
%%%%%%%% Pot. $\la x^2/(1 + g x^2)$ en M.Q.
%-----------------------------------------------------
\bibitem{BiDS73}{S.N. Biswas, K. Datta, R.P. Saxena, P.K. Srivastava,
and V.S. Varma},  1973
{\rm ``Eigenvalues of $\la\,x^{2m}$ anharmonic oscillators"},
{\sl J. Math. Phys.} {\bf 14}, 1190--1195.
%-----------------------------------------------------
\bibitem{MaL75}{P.M. Mathews and M. Lakshmanan},  1975
{\rm ``A quantum mechanically solvable nonpolynomial Lagrangian with
velocity-dependent interaction"},
{\sl Nuovo Cim.}  {\bf A 26},  299--315.
%-----------------------------------------------------
\bibitem{Mi78}{A.K. Mitra},  1978
{\rm ``On the interaction of the type $\la x^2/(1 + g x^2)$"},
{\sl J. Math. Phys.} {\bf 19}, 2018--2022.
%-----------------------------------------------------
\bibitem{Ka79}{R.S. Kaushal},  1979
{\rm ``Small $g$ and large $\la$ solution of the Schroedinger equation
for the interaction $\la x^2/(1 + g x^2)$"},
{\sl J. Phys. A} {\bf 19}, L253--L258.
%-----------------------------------------------------
\bibitem{BeB80}{N. Bessis and G. Bessis}, 1980
{\rm ``A note on the Schroedinger equation for the $x^2+\la x^2/(1 + g x^2)$
potential"},
{\sl J. Math. Phys.} {\bf 21}, 2780--2785.
%-----------------------------------------------------
\bibitem{Fl81}{N. Flessas}, 1981
{\rm ``On the Schroedinger equation for the $x^2+\la x^2/(1 + g x^2)$ 
interaction"},
{\sl Phys. Lett. A} {\bf 83}, 121--122.
%--------------------------------------------
%%%%%%%% Pot. no-lineal en M.Q. en n=3 dim
%-----------------------------------------------------
\bibitem{LaE75}{M. Lakshmanan and K. Eswaran},  1975
{\rm ``Quantum dynamics of a solvable nonlinear chiral model"},
{\sl J. Phys.} {\bf A 8}, 1658--1669.
%-----------------------------------------------------
\bibitem{BoV90}{S.K. Bose and N. Varma},  1990
{\rm ``Exact solution of the Schroedinger equation for the central 
nonpolynomial
potential $V(r)=r^2+\la r^2/(1 + g r^2)$ in two and three dimensions"},
{\sl Hadron. J.}  {\bf 13}, 47--56.
%-----------------------------------------------------
\bibitem{LiL70}{H.E. Lin, W.C. Lin, and R. Sugano},  1970
{\rm ``On velocity dependent potentials in quantum mechanics"},
{\sl Nucl. Phys.} {\bf B 16}, 431--449.
%-----------------------------------------------------
\bibitem{VeW71}{G. Velo and J. Wess},  1971
{\rm ``A solvable quantum mechanical model with nonlinear 
transformations laws"},
{\sl Nuovo Cim.}  {\bf A 1},  177--187.
%-----------------------------------------------------
%%%%%%%% Masa dep de la posici\'on
%-----------------------------------------------------
\bibitem{Le95}{J.M. L\'evy-Leblond}, 1995
{\rm ``Position-dependent effective mass and Galilean invariance"},
{\sl Phys. Rev} {\bf A 52}, 1485--1489.
%-----------------------------------------------------
%%%%%%%%
%-----------------------------------------------------
\bibitem{RaS02a}{M.F. Ra\~nada and M. Santander}, 2002
{\rm ``On some properties of the harmonic oscillator on spaces of
constant curvature"},
%%% (33rd Symposium on Mathematical Physics, Torun, 2001).
{\sl Rep. Math. Phys.} {\bf 49}, 335--343.
%-----------------------------------------------------
\bibitem{RaS02b}{M.F. Ra\~nada and M. Santander}, 2002
{\rm ``On the harmonic oscillator on the two-dimensional sphere $S^2$ and
the hyperbolic plane $H^2$"},
{\sl J. Math. Phys.} {\bf 43}, 431--451.
%-----------------------------------------------------
\bibitem{RaS03}{M.F. Ra\~nada and M. Santander}, 2003
{\rm ``On the harmonic oscillator on the two-dimensional sphere $S^2$
and the hyperbolic plane $H^2$. II"},
{\sl J. Math. Phys.} {\bf 44}, 2149--2167.
%-----------------------------------------------------
%%%%%%%% H.H.
%-----------------------------------------------------
\bibitem{Sa91}{W. Sarlet}, 1991
{\rm ``New aspects of integrability of generalized H\'enon--Heiles systems"},
{\sl J. Phys.} {\bf A 24}, 5245--5251.
%-----------------------------------------------------
\bibitem{RaGaCa}{V. Ravoson, L. Gavrilov, and R. Caboz}, 1993
{\rm ``Separability and Lax pairs for the H\'enon--Heiles system"},
{\sl J. Math. Phys.} {\bf 34}, 2385--2393.
%---------------------------------------------------
\bibitem{RWT96}{S. Rauch-Wojciechowski and A.V. Tsiganov}, 1996
{\rm ``Quasi-point separation of variables for the H\'enon--Heiles
system and a system with a quartic potential"},
{\sl J. Phys.} {\bf A 29}, 7769--7778.
%---------------------------------------------------
\bibitem{CaRa99}{J.F. Cari\~nena and M.F. Ra\~nada}, 1999
{\rm ``Helmholtz conditions and alternative Lagrangians:
Study of an integrable H\'enon--Heiles system"},
{\sl Internat. J. Theoret. Phys.} {\bf 38}, 2049--2061.
%-----------------------------------------------------
%%%%%%%% Osc. Arm. Fact compleja
%-----------------------------------------------------
\bibitem{Pe90}{A.M. Perelomov}, 1990
{\sl ``Integrable systems of classical mechanics and Lie algebras"},
  (Birk\-hauser, Ba\-sel).
%-----------------------------------------------------
\bibitem{CaMR02}{J.F. Cari\~nena, G. Marmo, and M.F. Ra\~nada}, 2002
{\rm ``Non-symplectic symmetries and bi-Hamiltonian structures
of the rational harmonic oscillator"},
{\sl J. Phys. A} {\bf 35}, L679--L686.
%--------------------------------------------
%%%%%%%% Pot. Super-separables
%-----------------------------------------------------
\bibitem{FrMS65}{T.I. Fris, V. Mandrosov, Y.A. Smorodinsky, M. Uhlir,
and P. Winternitz}, 1965
{\rm ``On higher symmetries in quantum mechanics"},
{\sl Phys. Lett.} {\bf 16}, 354--356.
%-----------------------------------------------------
\bibitem{Ev90}{N.W. Evans}, 1990
{\rm ``Superintegrability in classical mechanics"},
{\sl Phys. Rev.} {\bf A 41}, 5666--76.
%-----------------------------------------------------
\bibitem{GrPS95a}{C. Grosche, G.S. Pogosyan, and A.N. Sissakian}, 1995
{\rm ``Path integral discussion for Smorodinsky-Winternitz potentials"},
%%: Two and three--dimensional Euclidean spaces"},
{\sl Fortschr. Phys.} {\bf 43}, 453--521.
%-----------------------------------------------------
\bibitem{Ra97}{M.F. Ra\~nada}, 1997
{\rm ``Superintegrable $n=2$ systems, quadratic constants of motion,
and potentials of Drach"},
{\sl J. Math. Phys.} {\bf 38}, 4165--4178.
%-----------------------------------------------------
\bibitem{BeCR00}{S. Benenti, C. Chanu, and G. Rastelli}, 2000
{\rm ``The super-separability of the three-body inverse-square
Calogero system"},
{\sl J. Math. Phys.} {\bf 41}, 4654--4678.
%-----------------------------------------------------
\bibitem{TeTW01}{P. Tempesta, A.V. Turbiner, and P. Winternitz}, 2001
{\rm ``Exact solvability of superintegrable systems"},
  J. Math. Phys. {\bf 42}, 4248--4257.
%-----------------------------------------------------
\bibitem{BHS03}{A. Ballesteros, F.J. Herranz, M. Santander and T. 
Sanz-Gil},  2003
{\rm ``Maximal superintegrability on $N$-dimensional curved spaces"},
{\sl J. Phys.} {\bf A 36}, L93--L99.
%-----------------------------------------------------
%%%%%%%% Pot. + Espacios con curvatura
%-----------------------------------------------------
\bibitem{Hi79}{P.W. Higgs}, 1979
{\rm ``Dynamical symmetries in a spherical geometry I"},
{\sl J. Phys.} {\bf A 12}, 309--323.
%------------------------------------------------
\bibitem{Le79}{H.I. Leemon}, 1979
{\rm ``Dynamical symmetries in a spherical geometry II"},
{\sl J. Phys.} {\bf A 12}, 489--501.
%-----------------------------------------------------
\bibitem{GrPS95b}{C. Grosche, G.S. Pogosyan, and A.N. Sissakian}, 1995
{\rm ``Path integral discussion for Smorodinsky-Winternitz potentials II"},
%%%   Two and three--dimensional sphere,
{\sl Fortschr. Phys.} {\bf 43}, 523--563.
%-----------------------------------------------------
\bibitem{RaS99}{M.F. Ra\~nada and M. Santander}, 1999
{\rm ``Superintegrable systems on the two-dimensional sphere $S^2$
  and the hyperbolic plane $H^2$"},
{\sl J. Math. Phys.} {\bf 40}, 5026--5057.
%-----------------------------------------------------
\bibitem{Sl00}{J.J. Slawianowski}, 2000
{\rm ``Bertrand systems on spaces of constant sectional curvature"},
{\sl Rep. Math. Phys.} {\bf 46}, 429--460.
%-----------------------------------------------------
\bibitem{KaKP01}{E.G. Kalnins, J.M. Kress, G.S. Pogosyan, and W. Miller}, 2001
{\rm ``Completeness of superintegrability in two-dimensional
constant-curvature spaces"},
{\sl J. Phys.} {\bf A 34}, 4705--4720.
%-----------------------------------------------------
\bibitem{KaKW02}{E.G. Kalnins, J.M. Kress, and P. Winternitz}, 2002
{\rm ``Superintegrability in a two-dimensional space of nonconstant
curvature"},
{\sl J. Math. Phys.} {\bf 43}, 970--983.
%-----------------------------------------------------
\bibitem{KaMP02}{E.G. Kalnins, W. Miller, and G.S. Pogosyan}, 2002
{\rm ``The Coulomb-oscillator relation on $n$-dimensional spheres
and hyperboloids"},
{\sl Phys. of Atomic Nuclei} {\bf 65}, 1119--1127.
%-----------------------------------------------------
\bibitem{RSSBanach}{M.F. Ra\~nada, T. Sanz-Gil, and M. Santander}, 2002
{\rm ``Superintegrable potentials and superposition of Higgs oscillators
on the sphere $S\sp 2$"}  in  {\sl Classical and quantum integrability},
Banach Center Publ. {\bf 59}, 243--255 (Polish Acad. Sci., Warsaw).
%-----------------------------------------------------
%%%%%%%% Pot. de Scarf
%-----------------------------------------------------
\bibitem{Sc58}{F.L. Scarf}, 1958
{\rm ``New soluble energy band problem"},
{\sl Phys. Rev.}  {\bf 112}, 1137--1140.
%-----------------------------------------------------
\bibitem{DDS92}{R. De, R. Dutt, and U. Sukhatme}, 1992
{\rm ``Path-integral solutions for shape-invariance potentials using point
canonical transformations"},
{\sl Phys. Rev.}  {\bf A 46}, 6869--6880.
%-----------------------------------------------------
\bibitem{Gr93}{C. Grosche}, 1993
{\rm ``Path integral discussion for Scarf-like potentials"},
{\sl Nuovo Cim.}  {\bf B 108},  1365--1376.
%-----------------------------------------------------
%%%%%%%% Alg. cuadraticas
%-----------------------------------------------------
\bibitem{Dask01}{C. Daskaloyannis}, 2001
{\rm ``Quadratic Poisson algebras of two-dimensional classical
superintegrable systems and quadratic associative algebras of quantum
superintegrable systems"},
{\sl J. Math. Phys.}  {\bf 42},  1100--1119.
%------------------------------------------------




%-------------------------------------------------------------------
\end{thebibliography}
\end{document}